\documentclass[12pt]{article}
\usepackage{graphics}
\usepackage{amsthm}
\usepackage{amssymb}
\usepackage{amsmath}
\usepackage{amsfonts}
\usepackage{epic}
\usepackage{epsfig}
\usepackage{amscd}

\def\be{\begin{equation}}
\def\ee{\end{equation}}
\begin{document}

\begin{titlepage}
\begin{flushright}
UCB-PTH-08/12\\
\end{flushright}
%%%%%%%%%%%%%%%%%%%%%%%%%%%%%%%%%%%%%%%%%%%%%%%%%%%%%%%%%%%%%%%%%%%%%%%%
\begin{center}
\noindent{{\Large{Langlands duality
in Liouville-$H^3_+$ WZNW correspondence}}}

\vskip 1.00cm

\smallskip
\smallskip
\smallskip
\smallskip

\smallskip
\smallskip
\noindent{\large{Gaston Giribet $^1$, Yu Nakayama $^2$, Lorena Nicol\'as $^3$}}

\smallskip
\smallskip

\end{center}
\smallskip
\centerline{$^1$ Department of Physics, Universidad de Buenos Aires and CONICET}
\centerline{{\it Ciudad Universitaria, Pabell\'on I, 1428. Buenos Aires, Argentina.}}
\smallskip
\smallskip
\smallskip
\centerline{$^2$ Berkeley Center for Theoretical Physics and Department of Physics}
\centerline{{\it University of California, Berkeley, California 94720-7300.}}
\smallskip
\smallskip
\smallskip
\centerline{$^3$ Instituto de Astronom\'{\i}a y F\'{\i}sica del Espacio, CONICET}
\centerline{{\it Ciudad Universitaria, C.C. 67 Suc. 28, 1428, Buenos Aires, Argentina.}}
\smallskip

\smallskip

\vskip 3.00cm

\begin{abstract}
We show a physical realization of the Langlands duality in correlation functions of $H_3^+$ WZNW model.
We derive a dual version of the Stoyanovky-Riabult-Teschner (SRT) formula that relates the correlation function
of the $H_3^+$ WZNW and the dual Liouville theory to investigate the level duality $k-2 \to (k-2)^{-1}$ in the
WZNW correlation functions. Then, we show that such a dual version of the $H_3^+ -$Liouville relation can be
interpreted as a particular case of a biparametric family of non-rational CFTs based on the Liouville correlation
functions, which was recently proposed by Ribault. We study symmetries of these new non-rational CFTs and
compute correlation functions explicitly by using the free field realization to see how a generalized Langlands
duality manifests itself in this framework. Finally, we suggest an interpretation of the SRT formula as realizing the
Drinfeld-Sokolov Hamiltonian reduction. Again, the Hamiltonian reduction reveals the Langlands duality in the
$H_3^+$ WZNW model. Our new identity for the correlation functions of $H_3^+$ WZNW model may yield a
first step to understand quantum geometric Langlands correspondence yet to be formulated mathematically.

\end{abstract}

\end{titlepage}
%%%%%%%%%%%%%%%%%%%%%%%%%%%%%%%%%%%%%%%%%%%%%%%%%%%%%%%%%%%%%%%%%%%%%%%%

%\newpage

%\tableofcontents

\newpage

%%%%%%%%%%%%%%%%%%%%%%%%%%%%%%%%%%%%%%%%%%%%%%%%%%%%%%%%%%%%%%%%%%%%%%

\section{Introduction}

Two-dimensional non-rational conformal field theories (CFTs) have many
applications both in physics and mathematics, from quantum (stringy) black
hole in physics to the geometric Langlands program in mathematics. Most of
what we currently know about these theories, however, is based on our understanding of
Liouville field theory (LFT) \cite{Yu}. In fact, LFT is by far the best
understood theory among non-rational CFTs, which turns out to be the
prototypical model to establish their exact quantization. A clear example is
the $H_{3}^{+}=SL(2,\mathbb{C})/SU(2)$ Wess-Zumino-Novikov-Witten (WZNW)
theory, whose structure was actually understood by resorting to the analogy
with LFT \cite{T1,T4,Andreev}.

The story took a new direction three years ago when S. Ribault and J.
Teschner showed that the relation between LFT and WZNW model could be pushed
forward, beyond the level of a mere \textit{analogy}, to the level of 
\textit{correspondence} in correlation functions. In \cite{RT}, they proved
that arbitrary correlation functions of the $H_{3}^{+}\ $WZNW model admit
simple expressions in terms of correlation functions of LFT. More precisely,
any $n$-point function of the $H_{3}^{+}$ WZNW theory on the topology of the
sphere can be written in terms of a $2n-2$-point functions of LFT. This
correspondence between observables of these two non-rational CFTs follows
from a previous result of A. Stoyanovky, who proved in \cite{S} a surprising
functional relation between solutions to the Knizhnik-Zamolodchikov (KZ)
equation and to the Belavin-Polyakov-Zamolodchikov (BPZ) equation. In this
paper, we refer to the formula that connects WZNW correlation functions and
Liouville correlation functions as Stoyanovsky-Ribault-Teschner (SRT)
formula.

The primary aim of this paper is to further investigate the SRT $H_{3}^{+}-$%
Liouville correspondence and its generalizations, especially in order to
understand the Langlands level duality in correlation functions of $H_{3}^+$
WZNW model and study its physical applications. We, thus, begin with the
review of recent development in this direction.

\subsection{The $H_{3}^{+}-$Liouville correspondence}

The $H_{3}^{+}-$Liouville correspondence has several interesting
applications in string theory. For example, it can be straightforwardly
adapted to describe the $SL(2,\mathbb{R})_{k}/U(1)$ coset model, so that
string amplitudes in the two-dimensional black hole background can be
described by Liouville correlation functions \cite{F}. This correspondence
is also relevant to study string theory in three-dimensional Anti-de Sitter
space (AdS$_{3}$), the dynamics of inhomogeneous tachyon condensation in
closed string theory, the six-dimensional little string theory, and many
other scenarios (see \cite{Taka,NN,Taka2,Niar,G2005} and references
therein). Some of these applications were investigated in \cite{GN}, where
it was pointed out that in order to fully describe the tree-level string
amplitudes in AdS$_{3}$, the result of \cite{RT} needed to be generalized to
include the spectral flowed sectors\footnote{%
In this paper, we do not make a clear distinction between the $SL(2,\mathbb{R%
})$ WZNW model and $H_{3}^{+}$ model because we assume that the analytic
continuation of correlation functions describes the former from the latter.
See for instance \cite{MO3} for a similar treatment.} of $SL(2,\mathbb{R}%
)_{k}$. In \cite{R2}, S. Ribault achieved to incorporate such spectral
flowed sectors by extending the results of \cite{RT}. The key point was to
generalize the KZ equation to the case of WZNW correlation functions that
involve spectral flowed fields. In particular, it was shown that if in a
given WZNW correlation function the conservation of the spectral flow number
is violated in $\Delta \omega $ units, then such a correlation function can
always be written in terms of a $2n-2-\Delta \omega $-point correlation
function of LFT. This new correspondence increased the set of WZNW
correlation functions that admit a representation in terms of LFT (see
formula (\ref{YYYYT}) of Appendix B).

After the formulation of the SRT $H_{3}^{+}-$Liouville correspondence on the
worldsheet sphere, further generalizations were accomplished. First, its
extension to the case of worldsheet geometry with boundaries was worked out
in \cite{R2005,HR,R2007,FR}, which can be regarded as a worldsheet
description of the D-brane in the string theory context.

The second generalization was the extension to the case of higher genus
correlation functions: in \cite{HS} Y. Hikida and V. Schomerus proved that
any $n$-point correlation functions of the $H_{3}^{+}$ WZNW model at genus $%
g $ can be written in terms of $2n+2g-2$-point functions of LFT. This higher
genus generalization was done by employing a path integral derivation of the 
$H_{3}^{+}-$Liouville correspondence (see also \cite{HS2}).

Very recently, following the path integral approach of \cite{HS}, S. Ribault
proposed a novel generalization of $H_{3}^{+}-$Liouville correspondence,
arguing that LFT may provide a representation of observables of a wider set
of CFTs \cite{R}. According to this proposal, SRT $H_{3}^{+}-$Liouville
correspondence could be merely a particular example of a more general
correspondence. The statement \cite{R} is that $2n-2$-point correlation
functions of LFT on the sphere can be regarded as generators of $n$-point
correlation functions of a biparametric family of non-rational CFTs. Each
member of this family of theories is characterized by two continuous
parameters, $b$ and $m$, and the parameterization is such that LFT
corresponds to the particular case $m=0$, having central charge $%
c_{L}=1+6(b+b^{-1})^{2}$.

We here make a following preliminary observation, on which we elaborate more
in this paper. On one hand, among members of the above biparametric family, the $%
H_{3}^{+}$ WZNW model corresponds to the case $m=1$, where the WZNW level is
given by $k=b^{-2}+2$ and its central charge by $c_{SL(2)}=3+6b^{2}$. On the
other hand, as we will show, the case $m=b^{2}$ also corresponds to the $%
H_{3}^{+}$ WZNW theory whose central charge $c_{SL(2)}=3+6b^{-2}$, but with
level $k=b^{2}+2$. This implies that the $H_{3}^{+}$ model is represented by
two curves in the space of parameters ($m,b$) of \cite{R}. Fixing the level $%
k$ then corresponds to fixing a point on each curve, where the one curve
turns out to be related to the other by the Langlands level duality $%
k-2\rightarrow (k-2)^{-1}$. In this paper, we try to reveal the
manifestation of Langlands duality in the $H_{3}^{+}$ WZNW model, and we
will also argue that this could be seen as an example the more general
duality. Specifically, more members of the biparametric family of CFTs
proposed in \cite{R} could actually appear twice in the space of parameters
defined by the ($m,b$) plane. This idea is suggested by the structure of the
conformal Ward identities and it appears naturally when discussing the
current algebra that generates the symmetries of the theories.

\subsection{Langlands duality and WZNW\ theory}

The relation between the $H_{3}^{+}$ WZNW model (or $SL(2,\mathbb{R})_{k}$
WZNW model) and LFT reveals its significance not only in physics, but also
in mathematics. Long before the advent of the SRT relation, the connection
between these two CFTs had been studied within the context of the geometric
Langlands correspondence\footnote{%
In physics literature, the terminology \textquotedblleft Langlands duality"
is often used to refer to the geometric Langlands correspondence \cite%
{Informal}\ we are mentioning here. However, we would like to save the
terminology \textquotedblleft Langlands duality" to refer to a more specific
duality; namely the duality under level transformation $k-2\rightarrow
(k-2)^{-1}$ of the $\hat{sl}(2)_{k}$ current algebra. The reason why this
duality is referred to as \textquotedblleft Langlands duality" is that it
corresponds to the $\mathcal{W}$-algebra (Virasoro) duality isomorphism
which plays an important role in the geometric Langlands correspondence (see
section 8 of Ref. \cite{Frenkel:2005pa} and section 6 of Ref. \cite{BP}\ for
discussions).} (see e.g. \cite{Frenkel:2005pa, Frenkel:2006nm, Tan:2007ej,
Tan:2008ak}). It turns out that CFTs with affine Kac-Moody symmetry (let us
call the corresponding algebra $\hat{g}$) give a natural realization of the
geometric Langlands program, as they provide a natural way of realizing the
so-called Hecke eigensheaves. In this context, Hecke eigensheaf is an object
closely related to the chiral conformal blocks of the $G_{k}$ WZNW model,
and they are $\mathcal{D}$-modules on the moduli space of the $G$-bundle on
the worldsheet that are attached to the $^{L}G$-bundle with holomorphic
connection on the worldsheet, being $^{L}G$ the Langlands dual of the Lie
group $G$ associated to the Lie algebra $g$; see \cite{Frenkel:2005pa}.

One of the simplest but still highly nontrivial examples of the geometric
Langlands correspondence appears in the case of $\hat{g}=sl(2)_{k}$ at the
critical level $k\rightarrow 2$ (i.e. where the level $k$ takes the value of
the Coxeter number). On one hand, we have the sheaf of coinvariants\footnote{%
They can be shown to be the Hecke eigensheaves.} (roughly speaking, $%
\mathcal{D}$-modules associated to the chiral correlation functions or
conformal blocks) of the $sl(2)_{k=2}$ over the moduli space of holomorphic $%
SL(2)$-bundle on the worldsheet. On the other hand, we have a flat
holomorphic $^{L}SL(2)$-bundle on the worldsheet, which generates a
classical Virasoro algebra as its Poisson structure (upon Hamiltonian
reduction). The Langlands correspondence predicts a correspondence between
these two notions.

To study this connection, one first investigates the $sl(2)_{k}$ current
algebra at its critical point $k\rightarrow 2$. First of all, the structure
of chiral correlation functions depends on the $SL(2)$-bundle by varying
background $SL(2)$ gauge field in the action. In the Langlands
correspondence, we identify this deformation of correlation functions with
another deformation induced by changing the center of the $sl(2)_{k=2}$
current algebra which leads to a center-dependent representation of primary
vertex operators. It turns out that the center of the affine algebra is
generated by the un-normalized Sugawara current which yields the classical
Virasoro algebra\footnote{%
Notice that at the critical point $k=2$ the Virasoro current (the
stress-tensor $T(z)$ with appropriate normalization) commutes with the
Kac-Moody currents $J^{a}(z)$.} corresponding to a holomorphic $%
^{L}SL(2)=SL(2)/\mathbb{Z}_{2}$ connection on the worldsheet with the
structure of an $^{L}sl(2)$-oper (i.e. modulo gauge equivalence by its Borel
subalgebra).

The geometric Langlands correspondence formulated on the $sl(2)_{k}$ current
algebra in this way has an intimate connection with the Drinfeld-Sokolov
(DS) Hamiltonian reduction, which reduces the $SL(2,\mathbb{R})_{k}$ WZNW
model to LFT, \cite{Bershadsky:1989mf, Feigin:1991wy}. In the framework of
geometric Langlands correspondence, the isomorphism between the center of
the $sl(2)_{k}$ current algebra at the critical level and the classical
Virasoro algebra can also be seen as a corollary of the quantum Hamiltonian
reduction, namely, the isomorphism between the representations of
Hamiltonian reduced chiral algebra and those of the Virasoro algebra at the
quantum level\footnote{%
By using the Wakimoto free field construction, one can show that the BRST
cohomology of the quantum Hamiltonian reduction is the center of the $%
sl(2)_{k}$ current algebra at the critical level, which completes the
argument (see for instance \cite{Bershadsky:1989mf}).}. Beyond the classical
correspondence, the quantum Hamiltonian reduction also suggests a further
mysterious duality (Langlands duality \cite{Feigin:1991wy}) for the $%
sl(2)_{k}$ current algebra with $k\neq 2$. The quantum version of
Hamiltonian reduction connects the $sl(2)_{k}$ current algebra of level $k$
and the Virasoro algebra with the central charge $c=1+6(b+b^{-1})$, where $%
b^{-2}=k-2$. Notice that this is actually the point where the Langlands
correspondence relates the $SL(2,\mathbb{R})_{k}$ WZNW model at the critical
level ($k\rightarrow 2$) with the classical LFT ($b\rightarrow \infty $).
The crucial observation is that the same Virasoro algebra is obtained with
the Langlands dual $sl(2)_{\tilde{k}}$ algebra at the dual level $\tilde{k}$%
, where $b^{2}=\tilde{k}-2$. Thus, under Hamiltonian reduction, the $%
sl(2)_{k}$ current algebra manifests the intriguing level duality $%
k-2\rightarrow \tilde{k}-2=(k-2)^{-1}$. From the viewpoint the Virasoro
algebra the duality is nothing but the Liouville self-duality under $%
b\rightarrow b^{-1}$. However, from the viewpoint of the $sl(2)_{k}$ current
algebra, it is quite mysterious: not only it relates the strongly coupled
system with the weakly coupled system, but it also changes the central
charge as $c_{_{SL(2)}}=3+6b^{2}\rightarrow 3+6b^{-2}$.

In this paper, we attempt to shed more light on this level duality. More
precisely, we would like to investigate a possible connection among the
following three notions: the Langlands level duality $k-2\rightarrow \tilde{k%
}-2$, the SRT $H_{3}^{+}-$Liouville correspondence, and the DS Hamiltonian
reduction from the viewpoint of the correlation functions. To do this, we
will begin by deriving a generalization of the SRT relation that will
manifest the level duality. Indeed, one can reformulate the SRT formula by
essentially changing $b$ with $b^{-1}$ in the LFT side, and then we show
that it leads to a surprising relation between the correlation functions in
the $H_{3}^{+}$ WZNW theory with level $k$ and those with the dual level $%
\tilde{k}$. Such a duality relation at finite values of $k$ is actually
envisaged also by mathematicians, as somehow it encodes the quantum version
of the geometric Langlands correspondence. Mathematical understanding of the
Langlands correspondence at the off-critical level is under lively
investigation (e.g. \cite{Stoyanovsky:2006mj}, see also \cite{Bakas:2005sd,
Bakas2} from a physical account). In this sense, our dual version of SRT
formula can be seen as a physical intuition of quantum Langlands
correspondence yet to be formulated mathematically at the level of the full
correlation functions\footnote{%
We emphasize that our approach only gives a relation between the full
correlation functions and not between the chiral correlation functions.}.

In \cite{HS}, Hikida and Schomerus discussed the relevance of the SRT $%
H_{3}^{+}-$Liouville correspondence in the context of classical geometric
Langlands correspondence on higher genus curves as it precisely gives the
appropriate basis of the WZNW conformal blocks that can be expressed
directly in terms of the conformal blocks of the Virasoro algebra
(corresponding to the same Virasoro algebra obtained through DS Hamiltonian
reduction). Our version of the quantum Langlands duality for the correlation
functions can be straightforwardly generalized to higher genus
correspondence by virtue of their results, so it will give more insight
about the quantum Langlands duality on higher genus.

Before closing the introduction, a final word about the notation is in
order. As mentioned, understanding of the Langlands level duality beyond the
critical value is of significance both in physics and mathematics.
Nevertheless, our motivations are entirely based on physical grounds, and
consequently, our discussion will be in the language usually employed within
the physics context.

\subsection{Overview}

The rest of the paper is organized as follows. In section 2, we derive the
dual version of the $H_{3}^{+}-$Liouville correspondence formula of \cite{RT}%
. We do this by using both the algebraic method and the path integral
approach. In section 3, we show that the dual version of the $H_{3}^{+}-$%
Liouville formula can be interpreted as a particular case of the Lagrangian
representation of a biparametric family of non-rational CFTs recently
proposed in \cite{R}. We also compute the correlation functions in these
CFTs by using the free field theory representations to give explicit
expression for the three-point functions. Through the discussion, a
generalization of the Langlands level duality will be manifested among these
new non-rational CFTs. In section 4, we study the DS Hamiltonian reduction
at the level of correlation functions and its relation to the SRT $%
H_{3}^{+}- $Liouville correspondence. The Hamiltonian reduction
interpretation of the (dual) SRT formula yields a surprising identity
realizing the quantum Langlands duality. We conclude in section 5 with some
remarks and open questions. In Appendix A, we collect some information of
special functions used in the main text. In Appendix B, we generalize our
discussion in the case of winding number violating correlation functions.

\section{Dual version of SRT formula}

Our first goal is to derive the dual version of the $H_{3}^{+}-$Liouville
correspondence of \cite{RT}. The formula relates $n$-point functions of $%
H_3^+$ WZNW model in the so-called $\mu$ basis (see section 2-2 for more
details) and $(2n-2)$-point functions of LFT. More precisely, we would like
to propose the following dual version of SRT formula\footnote{%
To avoid a confusion, we use the notation $\tilde{b}$ for the Liouville
exponent to emphasize we are discussing the dual SRT formula, while we
eventually identify $\tilde{b}$ with $b$ when we study the Langlands duality.%
} 
\begin{equation}
\left\langle \prod\nolimits_{i=1}^{n}\Phi _{j_{i}}(\mu
_{i}|z_{i})\right\rangle _{H_{3}^{+}}=\frac{\pi }{2\tilde{b}}(-\pi
)^{n}\delta ^{(2)}\left( \sum\nolimits_{i=1}^{n}\mu _{i}\right) |\Theta
_{n}|^{2}\left\langle \prod\nolimits_{i=1}^{n}V_{\alpha
_{i}}(z_{i})\prod\nolimits_{t=1}^{n-2}V_{-\frac{\tilde{b}}{2}%
}(y_{t})\right\rangle _{LFT}\ ,  \label{UPP}
\end{equation}%
where the correlation function on the right hand side corresponds to a $2n-2$%
-point function of LFT, which involves $2n-2$ exponential primary fields $%
V_{\alpha }(z)=e^{\sqrt{2}\alpha \varphi (z)}$. The central charge of LFT is
given in terms of $\tilde{b}$ as $c_{L}=1+6Q^{2}$, $Q=\tilde{b}+\tilde{b}%
^{-1}$. We define LFT by the classical action 
\begin{equation*}
S_{LFT}=\frac{1}{2\pi }\int d^{2}z\left(\partial \varphi \bar{\partial}%
\varphi +2\pi \mu _{L}e^{\sqrt{2}\tilde{b}\varphi }\right) \ .
\end{equation*}

The interpolating function $\Theta _{n}$ is given by 
\begin{equation}
\Theta _{n}(z_{1},\cdots ,z_{n}|y_{1},\cdots ,y_{n-2}|u)=\frac{%
u\prod_{r<s\leq n}(z_{r}-z_{s})^{\frac{\tilde{b}^{2}}{2}}\prod_{t<l\leq
n-2}(y_{t}-y_{l})^{\frac{\tilde{b}^{2}}{2}}}{\prod_{r=1}^{n}%
\prod_{t=1}^{n-2}(z_{r}-y_{t})^{\frac{\tilde{b}^{2}}{2}}}\ ,  \label{TTita}
\end{equation}%
where $y_{i}$ are related to $\mu _{i}$ and $u$ (so-called Sklyanin's
separation of variables) as follows: 
\begin{equation}
u=\sum\nolimits_{i=1}^{n}\mu _{i}z_{i},\qquad \sum_{i=1}^{n}\frac{\mu _{i}}{%
t-z_{i}}=u\frac{\prod_{j=1}^{n-2}(t-y_{j})}{\prod_{i=1}^{n}(t-z_{i})}\ .
\label{Skilian}
\end{equation}%
The Liouville momenta $\alpha _{i}$ is related to the $SL(2,\mathbb{R})$%
-spin variables $j_{i}$ as 
\begin{equation}
\alpha _{i}=\tilde{b}^{-1}\left( j_{i}+1\right) +\tilde{b}/2\ ,  \label{PHA}
\end{equation}%
while the Liouville parameter $\tilde{b}$ is related to the WZNW level $k$ by%
\footnote{%
Note that in the original formula, we have $b^{2}=(k-2)^{-2}$.} 
\begin{equation}
\tilde{b}^{2}=k-2\ ,  \label{loimportante}
\end{equation}%
which implies the relation between conformal dimensions as $\Delta _{\alpha
}+\Delta _{-\tilde{b}/2}+\tilde{b}^{2}/2=\Delta _{\alpha }-k/4=-\tilde{b}%
^{-2}j(j+1)=\Delta _{j}$.

Expression (\ref{UPP}) represents a dual version of the SRT formula, as it
was presented in \cite{RT}. In fact, the original version of the formula in 
\cite{RT} is obtained from (\ref{UPP}) by replacing $\tilde{b}\rightarrow
b^{-1}$. This is actually the key point here: the fact that such a dual
expression exists implies that the formula (\ref{UPP}) and the one in \cite%
{RT}, considered together, induce the duality under ${b}^{-2}=k-2\rightarrow
(k-2)^{-1}$ at the level of WZNW correlation functions, provided the
self-dually under $b\rightarrow $ $\tilde{b}$ of Liouville theory holds and
a suitable transformation of the $SL(2,R)$-spin variables as\textbf{\ }$%
b(j+1+b^{-2}/2)\rightarrow \tilde{b}(j_{i}+1+\tilde{b}^{-2}/2)$ is
introduced. We return to this point in section 4.

Now, let us prove (\ref{UPP}) by reviewing the analysis of \cite{RT} and 
\cite{HS}.

\subsection{Knizhnik-Zamolodchikov equation}

Let us begin with the relation between reflection coefficients of both LFT\
and WZNW model. First, consider the Liouville two-point function 
\begin{equation}
R^{L}(\alpha )=-(\pi \mu _{L}\gamma (\tilde{b}^{2}))^{\frac{Q-2\alpha }{%
\tilde{b}}}\frac{\Gamma (1+\tilde{b}(2\alpha -Q))}{\Gamma (1-\tilde{b}%
(2\alpha -Q))}\frac{\Gamma (1+\tilde{b}^{-1}(2\alpha -Q))}{\Gamma (1-\tilde{b%
}^{-1}(2\alpha -Q)}\ .  \label{rlaplfa}
\end{equation}%
and the $SL(2,\mathbb{R})_{k}$ two-point function 
\begin{equation}
R^{H}(j)=-\left( \frac{\gamma (\frac{1}{k-2})}{\pi (k-2)}\right) ^{-2j-1}%
\frac{\Gamma (2j+1)}{\Gamma (-2j-1)}\frac{\Gamma (\frac{2j+1}{k-2})}{\Gamma
(-\frac{2j+1}{k-2})}\ ,  \label{rhjota}
\end{equation}%
where $\gamma (x)=\Gamma (x)/\Gamma (1-x)$. It is straightforward to verify
that reflection coefficients (\ref{rlaplfa}) and (\ref{rhjota}) are related
as 
\begin{equation}
R^{L}(\tilde{b}^{-1}(j+1)+\tilde{b}/2)=R^{H}(j)\ ,  \label{Rinq1}
\end{equation}%
as long as 
\begin{equation}
\left( \pi \mu _{L}\gamma (\tilde{b}^{2})\right) ^{\tilde{b}^{-2}}=\frac{%
\gamma (\frac{1}{k-2})}{\pi (k-2)}\ ,\qquad \text{or}\qquad \tilde{\mu}_{L}=%
\frac{1}{\pi ^{2}\tilde{b}^{2}}\ .  \label{Rinq2}
\end{equation}%
for the prefactor to match.\footnote{%
Recall the Liouville duality relation \cite{ZZ}: $\left( \pi \tilde{\mu}%
_{L}\gamma(b^{-2})\right) ^{b}=\left( \pi \mu _{L}\gamma (b^{2})\right)
^{1/b}$ . In \cite{RT} the convention $\mu =b^{2}/\pi ^{2}$ was used.} This
shows how the dual relation (\ref{UPP}) holds for the simple case of the
two-point function.

To go further, let us consider the Liouville four-point function,%
\begin{equation}
F^{L}(\alpha _{1},\alpha _{2},-\tilde{b}/2,\alpha _{3})=\left\langle
V_{\alpha _{1}}(z_{1})V_{\alpha _{2}}(z_{2})V_{-\frac{\tilde{b}}{2}%
}(y_{1})V_{\alpha _{3}}(z_{3})\right\rangle _{LFT} \ ,  \label{eqref}
\end{equation}%
which is given by 
\begin{eqnarray}
F^{L}(\alpha _{1},\alpha _{2},-\tilde{b}/2,\alpha _{3})
&=&|z_{32}|^{2(\Delta _{\alpha _{1}}-\Delta _{\alpha _{2}}-\Delta _{\alpha
_{3}}+\Delta _{-\tilde{b}/2})}|z_{31}|^{2(\Delta _{\alpha _{2}}-\Delta
_{\alpha _{1}}-\Delta _{\alpha _{3}}+\Delta _{-\tilde{b}/2})}\times  \notag
\\
&&\times |z_{21}|^{2(\Delta _{\alpha _{3}}-\Delta _{\alpha _{1}}-\Delta
_{\alpha _{2}}-\Delta _{-\tilde{b}/2})}|z_{3}-y_{1}|^{-4\Delta _{-\tilde{b}%
/2}}|1-z|^{2(j_{1}+1)+\tilde{b}^{2}}\times  \notag \\
&&\times \sum_{\eta =\pm }\left( \ |z|^{2(\Delta _{\alpha _{3}-\eta \tilde{ b%
}/{2}}-\Delta _{\alpha _{3}}+\Delta _{\eta \tilde{b}/{2}})}\tilde{C}_{\eta
}^{L}(\alpha _{3})C^{L}(\alpha _{2},\alpha _{1},\alpha _{3}-\eta \tilde{b}%
/2)\right. \ \times  \notag \\
&&\times \ \left. _{2}F_{1}(-j_{3}^{\eta }+j_{1}+j_{2}+1,-j_{3}^{\eta
}+j_{1}-j_{2},-2j_{3}^{\eta },z)\ \right)\ ,  \label{YYYY}
\end{eqnarray}%
where 
\begin{equation*}
z_{ab}=z_{a}-z_{b} \ ,\qquad z=\frac{(z_{1}-z_{2})(y_{1}-z_{3})}{%
(z_{1}-z_{3})(y_{1}-z_{2})}\ ,
\end{equation*}%
and $j^{-}=j$, $j^{+}=-j-1$. In (\ref{YYYY}), the function $C^{L}(\alpha
_{3},\alpha _{2},\alpha _{1})$ corresponds to Liouville structure constant 
\begin{equation*}
C^{L}(\alpha _{3},\alpha _{2},\alpha _{1})=(\pi \mu _{L}\gamma (\tilde{b}%
^{2})\tilde{b}^{2-2\tilde{b}^{2}})^{s}\frac{\Upsilon ^{\prime }(0)}{\Upsilon
(\alpha _{1}+\alpha _{2}+\alpha _{3}-Q)}\prod\nolimits_{i=1}^{3}\frac{%
\Upsilon (2\alpha _{i})}{\Upsilon (\alpha _{1}+\alpha _{2}+\alpha
_{3}-2\alpha _{i})}\ ,
\end{equation*}%
where $s=1+\tilde{b}^{-2}-\tilde{b}^{-1}(\alpha _{1}+\alpha _{2}+\alpha
_{3}) $. See Appendix A for the definition and properties of $\Upsilon (x)$.
The special structure constants $\tilde{C}_{\eta }^{L}(\alpha )$ in (\ref%
{YYYY}) are given by%
\begin{equation*}
\tilde{C}_{-}^{L}(\alpha )=(\pi \mu _{L}\gamma (\tilde{b}^{2}))^{\tilde{b}%
^{2}}\frac{\gamma (2\tilde{b}\alpha -1-\tilde{b}^{2})}{\gamma (2\tilde{b}%
\alpha )}\ ,\qquad \tilde{C}_{+}^{L}(\alpha )=1\ .
\end{equation*}

It is relatively easy to show that (\ref{eqref}) becomes%
\begin{eqnarray}
F^{L}(\alpha _{1},\alpha _{2},-\tilde{b}/2,\alpha _{3})
&=&|z_{32}|^{2(\Delta _{\alpha _{1}}-\Delta _{\alpha _{2}}-\Delta _{\alpha
_{3}}-\Delta _{-\tilde{b}/2})}|z_{31}|^{2(\Delta _{\alpha _{2}}-\Delta
_{\alpha _{1}}-\Delta _{\alpha _{3}}-\Delta _{-\tilde{b}/2})}\times  \notag
\\
&&\times |z_{21}|^{2(\Delta _{\alpha _{3}}-\Delta _{\alpha _{1}}-\Delta
_{\alpha _{2}}-\Delta _{-\tilde{b}/2})}|u|^{4\Delta _{-\tilde{b}/2}}|\mu
_{1}|^{\tilde{b}^{2}}|\mu _{2}|^{\tilde{b}^{2}}|\mu _{3}|^{\tilde{b}%
^{2}}\times  \notag \\
&&\times (-2\pi ^{2}\tilde{b})C^{H}(j_{3},j_{2},j_{1})D^{H}[j,\mu ] \ 
\label{estoq}
\end{eqnarray}%
with $C^{H}(j_{3},j_{2},j_{1})D^{H}[j,\mu ]$ being the $H_{3}^{+}$ WZNW
structure constants found in \cite{T1,T2,T3} written in terms of the
so-called $\mu $-basis introduced in \cite{RT}. Then, if one multiplies (\ref%
{estoq}) by $\Theta _{3}$, one finds the expected agreement, as stated in (%
\ref{UPP}).

To complete the proof, we have to study higher-point functions. For this
purpose, we first show the relation between the BPZ equation and the KZ
equation, corresponding to the right and the left hand side of (\ref{UPP})
respectively. The BPZ equation satisfied by LFT correlation functions that
involve degenerate field $V_{-\tilde{b}/2}$ is given by 
\begin{equation*}
\left[ \frac{1}{\tilde{b}^{2}}\frac{\partial ^{2}}{\partial y_{r}^{2}}%
+\sum_{s\neq r}\left( \frac{1}{y_{r}-y_{s}}\frac{\partial }{\partial y_{s}}+%
\frac{\Delta _{-\tilde{b}/{2}}}{(y_{r}-y_{s})^{2}}\right) +\sum_{s}\left( 
\frac{1}{y_{r}-z_{s}}\frac{\partial }{\partial z_{s}}+\frac{\Delta _{\alpha
_{s}}}{(y_{r}-z_{s})^{2}}\right) \right] \Omega _{2n-2}^{L}=0\ ,
\end{equation*}%
where $\Omega _{2n-2}^{L}$ denotes the Liouville correlation function
appearing in \eqref{UPP}. On the other hand, the Sklyanin separation of
variable yields the following form for the KZ equation 
\begin{equation*}
\left[ \frac{1}{\tilde{b}^{2}}\frac{\partial ^{2}}{\partial y_{a}^{2}}%
+\sum_{r=1}^{n}\frac{1}{y_{a}-z_{r}}\left( \frac{\partial }{\partial z_{r}}+%
\frac{\partial }{\partial y_{a}}\right) -\sum_{b\neq a}\frac{1}{y_{a}-y_{b}}%
\left( \frac{\partial }{\partial y_{a}}-\frac{\partial }{\partial y_{b}}%
\right) +\sum_{r=1}^{n}\frac{\Delta _{j_{r}}}{(y_{a}-z_{r})^{2}}\right]
\Omega _{n}^{H}=0\ ,
\end{equation*}%
where $\Omega _{n}^{H}$ denotes the $H_{3}^{+}$ correlation function in %
\eqref{UPP}.

Crucial observation is that these two equations agree with each other after
twisting by $\Theta_n$. Now, since the correlation functions of both
theories satisfy the same linear differential equation, one can show %
\eqref{UPP} by taking a particular limit $z_{12}\rightarrow 0$ and following
the same induction argument used in \cite{RT}.

In summary, the same argument employed in \cite{RT} leads to the derivation
of the dual relation (\ref{UPP}). Our dual formulation clearly implies the
close relation between the Liouville self duality under $\tilde{b} \to 
\tilde{b}^{-1} (=b)$ and the Langlands level duality under $k-2 \to
(k-2)^{-1}$ in $SL(2,\mathbb{R})_k$ WZNW correlation functions.
Generalization to the winding violating correlation functions and disk
one-point functions should be straightforward (see Appendix B).

We stress that although our dual formula looks as if it were a mere
rewriting of the original formula with the dual variable, it is not. It is
rather a consequence of the Liouville duality under $b\rightarrow b^{-1}$.
For example, we had to set the \textit{dual} cosmological constant to a
particular value and this relation is different from the original SRT
formula, whose origin will be clarified further when we discuss the path
integral derivation. Alternatively speaking, the inductive proof of the
equivalence between the original SRT formulation and the dual formulation
presented here can be thought of as a derivation of the self-duality of LFT
at the level of $n$-point correlation functions. Because the Liouville
self-duality is not trivial\footnote{%
For instance, it is known to break down in the spherical partition function 
\cite{Z}.} and not proven in general, our dual formula is indeed
non-trivial. We will see that the existence of such a dual formula leads to
a far-reaching consequence of the Langlands duality in the $H_{+}^{3}$%
-correlation functions.

In the next subsection, we present an alternative derivation of (\ref{UPP})
by using the path integral approach.

\subsection{Path integral derivation}

Dual SRT formula (\ref{UPP}) can be obtained also from the path integral
approach by using the so-called dual screening operator. The possibility was already
mentioned in \cite{HS}. The starting point is the $H_3^+$ WZNW model
represented by the free field action 
\begin{equation}
S_{0}=\frac{1}{2\pi }\int d^{2}z\left( \partial \phi \bar{\partial}\phi
+\beta \bar{\partial}\gamma +\bar{\beta}\partial \bar{\gamma}\right) \ ,
\label{dualSS}
\end{equation}%
where $\phi $ field has a background charge\footnote{%
We formulate the path integral on the flat Euclidean space. A careful
treatment of the curvature coupling can be found in \cite{HS}.} $\hat{Q}=1/%
\sqrt{k-2}=\tilde{b}^{-1}(=b)$, and by the addition of the $\mathit{dual}$
screening operator 
\begin{equation}
S_{s}=\frac{1}{2\pi }\int d^{2}z\left( -\beta \bar{\beta}\right) ^{\tilde{b}%
^{2}}e^{\sqrt{2}\tilde{b}\phi }\ .  \label{dualS}
\end{equation}

The $sl(2)_{k}$ vertex operators in the so-called $\mu $-basis can be
realized as%
\begin{equation*}
\Phi _{j}(\mu |z)=|\mu |^{2j+2}e^{\mu \gamma -\bar{\mu}\bar{\gamma}}e^{\sqrt{%
2}\tilde{b}^{-1}(j+1)\phi }\ ,
\end{equation*}%
whose conformal dimensions are $\Delta_j =-\tilde{b}^{-2}{j(j+1)}$.

To compute the left hand side of \eqref{UPP} and explicitly connect it with
the right hand side, we would like to evaluate the path integral 
\begin{equation*}
\left\langle \prod\nolimits_{i=1}^{n}\Phi _{j_{i}}(\mu
_{i}|z_{i})\right\rangle _{H_{3}^{+}}=\int \mathcal{D}\phi \mathcal{D}\gamma 
\mathcal{D}\bar{\gamma}\mathcal{D}\beta \mathcal{D}\bar{\beta}\
e^{-S_{0}-S_{s}}\prod\nolimits_{i=1}^{n}\Phi _{j_{i}}(\mu _{i}|z_{i})\ .
\end{equation*}%
The integration over field $\gamma $ (with a suitable contour modification)
yields the delta function constraint 
\begin{equation*}
\bar{\partial}\beta (w)=2\pi \sum_{i=1}^{n}\mu _{i}\delta (w-z_{i})\ ,
\end{equation*}%
or, equivalently, the integrated condition 
\begin{equation*}
\beta (w)=\sum_{i=1}^{n}\frac{\mu _{i}}{w-z_{i}}
\end{equation*}%
with $\sum_{i=1}^{n}\mu _{i}=0$. Then, we can introduce $y_{j}$ and $u$ such
that 
\begin{equation*}
\beta (w)=u\frac{\prod_{j=1}^{n-2}(w-y_{j})}{\prod_{i=1}^{n}(w-z_{i})}\ .
\end{equation*}

Integrating over field $\beta $ gives 
\begin{eqnarray*}
&&|u|^{2}\delta \left( \sum\nolimits_{i=1}^{n}\mu _{i}\right) \int \mathcal{D%
}\phi e^{-\frac{1}{2\pi }\int d^{2}w\left(\partial \phi \bar{\partial}\phi
+|u|^{2\tilde{b}^{2}}\left|
\prod_{t=1}^{n-2}(w-y_{t})\prod_{i=1}^{n}(w-z_{i})^{-1}\right| ^{2\tilde{b}%
^{2}}e^{\sqrt{2}\tilde{b}\phi }\right) }\times \\
&&\qquad \qquad \qquad \qquad \qquad \times \prod\nolimits_{i=1}^{n}|\mu
_{i}|^{2(j_{i}+1)}e^{\sqrt{2}\tilde{b}^{-1}(j_{i}+1)\phi }\ .
\end{eqnarray*}%
Now, to remove the prefactor in front of the interaction, we define the new
field 
\begin{equation*}
\varphi (w)=\phi (w)+\sqrt{2}\tilde{b}\log |u|^{2}+\sqrt{2}\tilde{b}\left(
\sum_{j=1}^{n-2}\log |w-y_{j}|^{2}-\sum_{i=1}^{n}\log |w-z_{i}|^{2}\right) \
.
\end{equation*}%
This yields the path integral representation 
\begin{eqnarray*}
\left\langle \prod\nolimits_{i=1}^{n}\Phi _{j_{i}}(\mu
_{i}|z_{i})\right\rangle _{H_{3}^{+}} &=&|\Theta _{n}|^{2}\delta \left(
\sum\nolimits_{i=1}^{n}\mu _{i}\right) \int \mathcal{D}\varphi e^{-\frac{1}{%
2\pi }\int d^{2}w\left( \frac{1}{2}\partial \varphi \bar{\partial}\varphi
+e^{\sqrt{2}\tilde{b}\varphi }\right) }\times \\
&&\times \prod\nolimits_{i=1}^{n}e^{(\sqrt{2}\tilde{b}^{-1}(j_{i}+1)+\frac{%
\sqrt{2}\tilde{b}}{2})\varphi (z_{i})}\prod\nolimits_{l=1}^{n-2}e^{-\frac{%
\sqrt{2}\tilde{b}}{{2}}\varphi (y_{l})}\ ,
\end{eqnarray*}%
where the background charge of $\varphi $ is given by $Q=\tilde{b}+\tilde{b}%
^{-1}$. The shift $\hat{Q}=\tilde{b}^{-1}\rightarrow Q=\tilde{b}+\tilde{b}%
^{-1}$ of the background charge could be understood by keeping track of the
curvature coupling as in \cite{HS}. Notice that the right hand side of the
equation above corresponds to the expected $2n-2$-point function of LFT. In
this way, we have obtained the path integral derivation of the dual SRT
formula.

In retrospect of the path integral derivation, we could interpret the dual
formula from the original variable $b$ instead of $\tilde{b}$. If we did
this, we would end up with the Liouville action with the \textit{dual}
Liouville interaction $e^{\sqrt{2}b^{-1}\phi }$. This is the reason why we
have to set the dual cosmological constant (and not the original
cosmological constant) to a particular value in \eqref{Rinq2}. This clearly
shows that the dual screening charge in $H_3^+$ model corresponds to the
dual Liouville interaction.

\section{A biparametric family of CFTs}

In this section, we will study relation (\ref{UPP}) in the context of the
generalization of SRT correspondence recently proposed in \cite{R}. We will
analyze the biparametric family of non-rational CFTs there to show that the
dual version of the SRT $H_{3}^{+}-$Liouville formula discussed in section 2
can be interpreted as a particular case of the theories described \cite{R}.

Let us consider the following quantity \cite{R}%
\begin{equation}
\Omega _{n}^{(m)}=\delta ^{(2)}\left( \sum\nolimits_{i=1}^{n}\mu _{i}\right)
|\Theta _{n}^{(m)}|^{2m^{2}}\left\langle \prod\nolimits_{i=1}^{n}V_{\alpha
_{i}}(z_{i})\prod\nolimits_{t=1}^{n-2}V_{-\frac{m}{2b}}(y_{t})\right\rangle
_{LFT} \ ,  \label{principal}
\end{equation}%
where $m$ and $b$ are two (real-valued) continuous parameters. The
coordinates $u,$ $z_{i},$ $y_{r}$ and $\mu _{i}$ are related through the
Sklyanin change of variable (\ref{Skilian}), and the function $\Theta
_{n}^{(m)}$ is defined by%
\begin{equation}
\Theta _{n}^{(m)}(z_{1},\cdots ,z_{n}|y_{1},\cdots ,y_{n-2}|u)=\frac{u^{(%
\frac{1}{m}+b^{-2}(\frac{1}{m}-1))}\prod_{r<s\leq n}\left(
z_{r}-z_{s}\right) ^{\frac{1}{2b^{2}}}\prod_{t<l\leq n-2}\left(
y_{t}-y_{l}\right) ^{\frac{1}{2b^{2}}}}{\prod_{r=1}^{n}\prod_{t=1}^{n-2}%
\left( z_{r}-y_{t}\right) ^{\frac{1}{2b^{2}}}} \ .  \label{ALPE}
\end{equation}%
In (\ref{principal}), as in (\ref{UPP}), the correlation function in the
right hand side corresponds to a $2n-2$-point function in LFT. This
correlation function involves $2n-2$ exponential primary fields $V_{\alpha
}(z)=e^{\sqrt{2}\alpha\varphi (z)}$, with $n-2$ of these fields having
momentum $\alpha =-m/2b$.

In \cite{R}, it was argued that $\Omega _{n}^{(m)}$ defined as in (\ref%
{principal}) could be interpreted as a correlation function of a certain
CFT, which is characterized by $m$ and $b$. This means that $\Omega
_{n}^{(m)}$ could be written as%
\begin{equation}
\Omega _{n}^{(m)}=\left\langle \prod\nolimits_{i=1}^{n}\Phi _{j_{i}}(\mu
_{i}|z_{i})\right\rangle _{CFT} \ ,  \label{CORR}
\end{equation}%
where $\Phi _{j}(\mu |z)$ would correspond to primary operators of a CFT.
This CFT is conjectured to exist, and it is considered \textquotedblleft
solvable" in the sense that its correlation functions are known provided the
LFT representation (\ref{principal}) is given. In turn, (\ref{principal}) is
thought of as a definition of a biparametric family of new non-rational
CFTs. The $H_{3}^{+}$ WZNW theory corresponds to the particular case $m=1$
as (\ref{principal}) reduces to the $H_{3}^{+}-$Liouville correspondence of 
\cite{RT}. On the other hand, LFT is obtained in the trivial case $m=0$. It
is worth noticing that the dual version of the SRT formula we derived in
section 2 is obtained at $m=b^{2}$ in (\ref{principal}). In fact, in this
case, the right hand side of (\ref{principal}) coincides with the right hand
side of (\ref{UPP}).

\subsection{Free field realization}

In \cite{R}, a Lagrangian representation of this family of CFTs was given by
generalizing the path integral approach of \cite{HS}. The Lagrangian for the
CFT of which (\ref{CORR}) are its correlation functions is given by the
action%
\begin{equation}
S[\lambda ]=\frac{1}{2\pi }\int d^{2}z\left( \partial \phi \bar{\partial}%
\phi +\beta \bar{\partial}\gamma +\bar{\beta}\partial \bar{\gamma}+\frac{%
Q_{m}}{2\sqrt{2}}R\phi +2\pi \lambda (-\beta \bar{\beta})^{m}e^{\sqrt{2}%
b\phi }\right)  \label{SW}
\end{equation}%
with the background charge $Q_{m}=b+b^{-1}(1-m)$. Here, $\lambda $
represents a real coupling constant whose specific value is controlled by
the zero mode of $\phi $. It is easy to verify that the interaction term $%
(-\beta \bar{\beta})^{m}e^{\sqrt{2}b\phi }$ has a conformal dimension $(1,1)$%
. Realization (\ref{SW}) is actually reminiscent of the Lagrangian
representation of the $SL(2,\mathbb{R})_{k}$ WZNW model. In fact, (\ref{SW})
does agree with the Wakimoto free field representation of the $SL(2,\mathbb{R%
})_{k}$ model in the particular case(s) $m=1$ (and $m=b^{2} (=\tilde{b}^2 =
k-2)$), where the WZNW level $k$ is given by $k=b^{-2}+2$ (resp. $k=b^{2}+2$%
).

Lagrangian realization (\ref{SW}) enabled us to study the symmetry algebra
underlying the solvable CFT \cite{R}, which is generated by the stress tensor%
\begin{equation}
T(z)=-\beta (z)\partial \gamma (z)-\frac{1}{2}(\partial \phi
(z))^{2}+(b+b^{-1}(1-m))\partial ^{2}\phi (z)  \label{ElT}
\end{equation}%
and the Borel subalgebra of the following representation of the affine
algebra $\hat{sl}(2)_{k}$%
\begin{eqnarray}
J^{+}(z) &=&\beta (z) \ ,  \label{A} \\
J^{-}(z) &=&\beta (z)\gamma ^{2}(z)-\sqrt{2}mb^{-1}\gamma (z)\partial \phi
(z)+(m^{2}b^{-2}+2)\partial \gamma (z) \ ,  \label{B} \\
J^{3}(z) &=&-\beta (z)\gamma (z)+\frac{1}{\sqrt{2}}mb^{-1}\partial \phi (z)
\ .  \label{C}
\end{eqnarray}

Here, as usual, fields $\beta $ and $\gamma $ form a commuting ghost system,
while field $\phi $ is a free boson with background charge $%
Q_{m}=(b+b^{-1}(1-m))$. These fields have non-vanishing propagators given by%
\begin{equation*}
\left\langle \beta (z)\gamma (w)\right\rangle =(z-w)^{-1} \ ,\qquad
\left\langle \phi (z,\bar{z})\phi (w,\bar{w})\right\rangle =-\log |z-w|^2 \ .
\end{equation*}

The central charge associated to stress tensor (\ref{ElT}) is given by $%
c=3+6Q_{m}^{2}$. It is worth noticing that interaction term $(-\beta \bar{%
\beta})^{m}e^{\sqrt{2}b\phi }$ in (\ref{SW}) commutes with (\ref{ElT}) for
any value of $m$ but it commutes with the currents (\ref{A})-(\ref{C}) only
for $m=1$ and $m=b^{2}$. In particular, the operator product expansion (OPE)
with $J^{-}(z)$ yields%
\begin{eqnarray*}
J_{(z)}^{-}\beta _{(w)}^{m}e^{\sqrt{2}b\phi (w)} &\sim &\frac{m}{(z-w)^{2}}%
\left( (m^{2}b^{-2}-m+1)\beta ^{m-1}e^{\sqrt{2}b\phi }+\right. \\
&&\left. +(z-w)(\sqrt{2}mb^{-1}\partial \phi \beta +(m-1)\partial \beta
)\beta ^{m-2}e^{\sqrt{2}b\phi }\right) +... \ ,
\end{eqnarray*}%
which in the cases $m=1$ and $m=b^{2}$ yields total derivatives%
\begin{eqnarray*}
J_{(z)}^{-}\beta _{(w)}e^{\sqrt{2}b\phi (w)} &\sim &b^{-2}\partial _{w}\frac{%
e^{\sqrt{2}b\phi (w)}}{(z-w)}+... \\
J_{(z)}^{-}\beta _{(w)}^{b^{2}}e^{\sqrt{2}b\phi (w)} &\sim &b^{+2}\partial
_{w}\frac{\beta _{(w)}^{b^{2}-1}e^{\sqrt{2}b\phi (w)}}{(z-w)}+...
\end{eqnarray*}%
respectively. As a result, for generic values of $m$ and $b$, the symmetries
of theory (\ref{SW}) turns out to be generated by the Virasoro current $T(z)$
and the subalgebra generated by $J^{+}(z)$ and $J^{3}(z)$.

Lagrangian realization (\ref{SW}) also provides the explicit form of the
primary operators $\Phi _{j}(\mu |z)$, which read%
\begin{equation*}
\Phi _{j}(\mu |z)=|\mu |^{2m(j+1)}e^{\mu \gamma (z)-\bar{\mu}\bar{\gamma}(%
\bar{z})}e^{\sqrt{2}b(j+1)\phi (z,\bar{z})} \ .
\end{equation*}%
These are Virasoro primary fields of dimension%
\begin{equation}
\Delta _{j}=-(j+1)(b^{2}j+m-1) \   \label{ConfDim}
\end{equation}%
with respect to the stress tensor (\ref{ElT}). Notice that momenta $j_{i}$
and Liouville momenta $\alpha _{i}$ in (\ref{principal}) are related by $%
\alpha _{i}=b(j_{i}+1+mb^{-2}/2)$.

It is natural to consider the following representation for the vertex
operators,%
\begin{equation}
\Phi _{j,p,\bar{p}}(z)\sim \gamma _{(z)}^{p-m(j+1)}\bar{\gamma}_{(\bar{z})}^{%
\bar{p}-m(j+1)}e^{\sqrt{2}b(j+1)\phi (z,\bar{z})}\ .  \label{Vjp}
\end{equation}%
Again, this is reminiscent of the Wakimoto free field representation of the $%
SL(2,\mathbb{R})_{k}$ WZNW model, and yields the relation 
\begin{equation}
\Phi _{j,p,\bar{p}}(z)=\frac{\Gamma (1+p-m(j+1))}{\Gamma (m(j+1)-\bar{p})}%
\int d^{2}\mu \ \mu ^{-p-1}\ \bar{\mu}^{-\bar{p}-1}\ \Phi _{j}(\mu |z) \ .
\label{Toau}
\end{equation}%
The relation between basis $\Phi _{j,p,\bar{p}}(z)$ and $\Phi _{j}(\mu |z)$
follows from the functional relations $\Gamma (n)\Gamma (1-n)=(-)^{n}\Gamma
(0)$ and $\int ds$\ $e^{-st}s^{x-1}=t^{-x}\Gamma (x)$. Operators (\ref{Vjp}%
)\ obey the following OPE with respect to the current algebra (\ref{A})-(\ref%
{C})%
\begin{eqnarray*}
J^{\pm }(z)\Phi _{j,p,\bar{p}}(w) &\sim &\frac{p_{i}\mp m(j_{i}+1)}{(z-w)}\
\Phi _{j,p\mp 1,\bar{p}}(w)+... \\
J^{3}(z)\Phi _{j,p,\bar{p}}(w) &\sim &\frac{-p_{i}}{(z-w)}\ \Phi _{j,p,\bar{p%
}}(w)+...
\end{eqnarray*}%
so that, in particular, these are Kac-Moody primaries under the Borel
subalgebra generated by $J^{+}(z)$ and $J^{3}(z)$, which are symmetry of the
system.

In the next subsection, we will compute three-point functions of vertex
operators (\ref{Vjp}) by employing the Lagrangian realization (\ref{SW}) for
a generic member of the biparametric family of CFTs, but, first, let us
discuss a particular case: note that if we specify $m=b^{2}(=\tilde{b}%
^{2}=k-2)$ in (\ref{ALPE}), and compare it with the definition (\ref{TTita}%
), we get the relation\footnote{%
In particular, in the case $m=1$ function $\Theta _{n}^{(1)}$ coincides with
the function $\Theta _{n}$ of \cite{RT}.}%
\begin{equation}
|\Theta _{n}^{(b^{2})}|^{2b^{4}}=|\Theta _{n}|^{2} \ .  \label{Aste}
\end{equation}%
\ This implies that $m=b^{2}$ in (\ref{SW}) yields an alternative
representation of the $H_{3}^{+}$ WZNW theory. We note that this
representation was the one employed in \cite{GN3} to explicitly compute WZNW
three-point functions. When $m=b^{2}$ the interaction term in (\ref{SW})
corresponds to the dual screening charge (\ref{dualS}), namely $S_{s}\sim
\int d^{2}z\beta ^{k-2}\bar{\beta}^{k-2}e^{\sqrt{2k-4}\phi }$, where the
relation with the WZNW level is now $m=b^{2}=k-2$. This has to be compared
with the case $m=1$ in (\ref{SW}), which corresponds to the standard
Wakimoto representation with the screening $S_{s}\sim \int d^{2}z\beta \bar{%
\beta}e^{\sqrt{\frac{2}{k-2}}\phi }$, with $b^{2}=(k-2)^{-1}$. Notice that
the relation between \thinspace the Liouville parameter $b$ and the WZNW\
level $k$ in each case is different; one is related to each other by $%
k-2\rightarrow (k-2)^{-1}$. This shows that in the framework of $H_{3}^{+}-$%
Liouville correspondence Langlands duality turns out to be induced by the
Liouville self-duality under $b\rightarrow b^{-1}$.

In summary, the dual version of the $H_{3}^{+}-$Liouville correspondence we
discussed in section 2 corresponds to a particular case of the Lagrangian
representation (\ref{SW}), that with $m=b^{2}=k-2$. This implies the $%
H_{3}^{+}$ WZNW theory turns out to be double-represented within the family
of CFTs proposed in \cite{R}. The $H_{3}^{+}$ model is represented by two
different curves in the space of parameters, and fixing the level $k$
corresponds to fixing a point on each curve. One curve is related to the
other by the level duality $k-2\rightarrow (k-2)^{-1}$, and this agrees with
free field realizations considered in the literature. The idea we would like
to suggest is that, presumably, this double-representation of CFTs within
the biparametric family of \cite{R} is a more general feature, and not only
happens to the $H_{3}^{+}$ WZNW theory. In fact, the structure of the
conformal Ward identities suggests that the CFT corresponding to the case $%
m=n$ (for a positive integer number $n\in \mathbb{Z}_{>0}$) coincides with
that corresponding to the case $m=nb^{2}$. Moreover, notice that function $%
|\Theta _{n}^{(m)}|^{2m^{2}}$ in (\ref{principal}) is such that the change $%
m\rightarrow mb^{2}$ can be always reinterpreted as the inversion $%
b\rightarrow b^{-1}$ but keeping $m$ fixed; and the same happens with the
auxiliary fields $V_{-m/2b}$ in the right hand side of (\ref{principal}).
Thus, assuming Liouville self-duality, one is led to conclude that both cases $%
m=n$ and $m=nb^{2}$ do correspond to the same CFT. It would be interesting
to explore this aspect, as it would yield a generalization of what Langlands
level duality is for the WZNW\ theory.

Before going into the explicit computation of correlation functions, we
would like to mention one open question about the CFTs described by (\ref{SW}%
). This is the question of identifying such CFTs. These theories likely
correspond to actual CFTs; but, which CFTs are those? We have just commented
that the particular case $m=b^{2}$ also corresponds to the $H_{3}^{+}$ WZNW;
however, analyzing in detail other particular cases seems to be a more
difficult problem. What we certainly know about the CFTs proposed in \cite{R}
is that they likely exist, and that a subset of their observables are given
by (\ref{principal}). Nevertheless, it remains a hard task to attempt a
classification, or to identify more particular cases. We could also ask
whether additional correlation functions other than those in (\ref{principal}%
)-(\ref{CORR}) are required to fully characterize the set of observables.
For instance, we know that this is actually the case for $m=1$ and $m=b^{2}$%
, where the spectral flowed sectors require a different amount of Liouville
fields on the right hand side of (\ref{principal}). Since spectral flow
symmetry is still an automorphism of the remnant algebra generated by $%
J^{3,-}(z)$ and $T(z)$, it is likely that a different amount of Liouville
insertions in (\ref{principal}) would also correspond to well-defined
correlation functions of the theories described by (\ref{SW}). This
certainly deserves further analysis. In the next subsection, we study the
explicit form of correlation functions to reveal some features of these
hypothetical CFTs.

\subsection{Correlation functions}

Let us compute correlation functions of the CFTs defined by (\ref{SW}). We
study the $n$-point correlation functions (\ref{CORR}) in the $p$-basis%
\begin{equation}
\Omega _{n}^{(m)}=\left\langle \prod\nolimits_{i=1}^{n}\Phi _{j_{i},p_i,\bar{%
p}_i}(z_{i})\right\rangle _{CFT} \ ,  \label{ENON}
\end{equation}%
which are defined by%
\begin{equation*}
\Omega _{n}^{(m)}=\int \mathcal{D}\phi \mathcal{D}\gamma \mathcal{D}\bar{%
\gamma}\mathcal{D}\beta \mathcal{D}\bar{\beta}\ e^{-S[\lambda
]}\prod\nolimits_{i=1}^{n}\gamma _{(z_{i})}^{p_{i}-m(j_{i}+1)}\bar{\gamma}_{(%
\bar{z}_{i})}^{\bar{p}_{i}-m(j_{i}+1)}e^{\sqrt{2}b(j_{i}+1)\phi (z_{i},\bar{z%
}_{i})} \ .
\end{equation*}

After integrating out the zero-modes, the correlation function can be
written as%
\begin{equation*}
\Omega _{n}^{(m)}=(-1)^{ms}\lambda ^{s}b^{-1}\Gamma (-s)\delta \left(
\sum\nolimits_{i=1}^{n}j_{i}+n+s-1-b^{-2}(1-m)\right) \times
\end{equation*}%
\begin{eqnarray}
&&\times \int \prod\nolimits_{r=1}^{s}d^{2}w_{r}\int \tilde{\mathcal{D}}\phi 
\mathcal{D}\gamma \mathcal{D}\bar{\gamma}\mathcal{D}\beta \mathcal{D}\bar{%
\beta}\ e^{-S[\lambda =0]}\prod\nolimits_{r=1}^{s}\beta _{(w_{r})}^{m}e^{%
\sqrt{2}b\phi (w_{r})}\times  \notag \\
&&\times \prod\nolimits_{i=1}^{n}\gamma _{(z_{i})}^{p_{i}-m(j_{i}+1)}e^{%
\sqrt{2}b(j_{i}+1)\phi (z_{i})}\times c.c. \ ,  \label{IIII}
\end{eqnarray}%
where $c.c.$ refers to the complex conjugate contribution, and the path
integral measure $\tilde{\mathcal{D}}\phi$ in the second line has to be
understood as excluding the zero mode. The integration over the zero mode of 
$\phi $ yields the first line in (\ref{IIII}), implying the condition%
\begin{equation}
\sum_{r=1}^{n}j_{r}+n+s=1+b^{-2}(1-m) \ ,  \label{ZUNA}
\end{equation}%
which, combined with the Riemann-Roch theorem, yields%
\begin{equation}
\sum_{r=1}^{n}\bar{p}_{r}=\sum_{r=1}^{n}p_{r}=(mb^{-2}-1)(1-m)\ .
\label{ZDOS}
\end{equation}

As usually happens in non-rational theories, expression (\ref{IIII}) has to
be understood just formally: since the kinematical configurations in (\ref%
{ZUNA}) can yield non-integer values $s$ of screening charges, the integrals
and products in (\ref{IIII}) (and consequently the Wick contractions, see
also \cite{Rasmussen1}) generally do not seem to be well-defined.
Nevertheless, these are usual features in non-rational CFTs and fortunately
the analytic continuation of such expressions is under control. Besides, for
positive integer values of $s$, the overall factor $\Gamma (-s)$ in (\ref%
{IIII}) diverges. This divergence is interpreted as due to the
non-compactness of the target space as in the case of LFT \cite{diFK}.

Performing the Wick contractions in (\ref{IIII}), we find the following
expression%
\begin{eqnarray*}
\Omega _{n}^{(m)} &=&(-1)^{ms}\lambda ^{s}b^{-1}\Gamma (-s)\delta \left(
\sum\nolimits_{i=1}^{n}j_{i}+n+s-1-b^{-2}(1-m)\right) \times \\
&&\times \prod\nolimits_{a<b\leq n}\left| z_{ab}\right|
^{-4b^{2}(j_{a}+1)(j_{b}+1)}\int
\prod\nolimits_{r=1}^{s}d^{2}w_{r}\prod\nolimits_{r<t\leq s}\left|
w_{r}-w_{t}\right| ^{-4b^{2}}\times \\
&&\times \prod\nolimits_{r=1}^{s}\prod\nolimits_{a=1}^{n}\left|
z_{a}-w_{r}\right| ^{-4b^{2}(j_{a}+1)}\times
\sum_{(i,r)}\prod\nolimits_{r=1}^{s}\prod%
\nolimits_{i=1}^{n}w_{i}^{r}(z_{i}-w_{r})^{-c_{i}^{r}}\times c.c. \ ,
\end{eqnarray*}%
where the sum $\sum_{(i,r)}$ runs over the different ways of choosing pairs (%
$i,r$) among $ms-1$, and with $i=1,...n$ while $r=1,...s$. Such
contributions correspond to the different combinations when performing the
Wick contractions of $\beta $-$\gamma $ fields, and coefficients $w_{i}^{r}$
are the multiplicity factors that count the different ways of contracting $%
c_{i}^{r}$ fields, piking those up \ among $r\leq s$ and $i\leq $ $%
p_{i}-m(j_{i}+1)$.

Now, let us focus on the three-point functions, which generically take the
form%
\begin{equation*}
\left\langle \prod\nolimits_{i=1}^{n=3}\Phi _{j_{i},p_{i},\bar{p}%
_{i}}(z_{i})\right\rangle _{CFT}=|z_{12}|^{2(\Delta _{j_{3}}-\Delta
_{j_{1}}-\Delta _{j_{2}})}|z_{13}|^{2(\Delta _{j_{2}}-\Delta _{j_{3}}-\Delta
_{j_{1}})}|z_{23}|^{2(\Delta _{j_{1}}-\Delta _{j_{2}}-\Delta
_{j_{3}})}C_{(j_{1},j_{2},j_{3}|p_{1},p_{2},p_{3})}^{(m)} \ ,
\end{equation*}%
where $C_{(j_{1},j_{2},j_{3}|p_{1},p_{2},p_{3})}^{(m)}$ represent the
structure constants. Here, we will concentrate on the case $p_{2}=\bar{p}%
_{2}=m(1+j_{2})$ because the computation becomes drastically simpler: the
Wick contraction of the $\gamma $-$\beta $ system can be carried out without
resorting to abstruse combinatorics. Nevertheless, the computation for the
generic case can be done by a suitable adaptation of the results of \cite%
{Satoh}. See for instance Eq. (2.15) of \cite{Satoh}, and cf. Eq. (3.20) in 
\cite{BB}. We will not address the case $p_{2}\neq m(1+j_{2})$ here.

Let us compute $C_{(j_{1},j_{2},j_{3}|p_{1},m(1+j_{2}),p_{3})}^{(m)}$. To do
this, first we need to compute the Wick contractions 
\begin{equation*}
W_{(j_{1},j_{2},j_{3}|p_{1},m(1+j_{2}),p_{3})}^{(m)}=\int
\prod\nolimits_{r=1}^{s}d^{2}w_{r}\left\langle e^{\sqrt{2}b(j_{1}+1)\phi
(0)}e^{\sqrt{2}b(j_{2}+1)\phi (1)}e^{\sqrt{2}b(j_{3}+1)\phi (\infty
)}\prod\nolimits_{r=1}^{s}e^{\sqrt{2}b\phi (w_{r},\bar{w}_{r})}\right\rangle
_{\lambda =0}
\end{equation*}%
\begin{equation}
\times \left\langle \gamma _{(0)}^{p_{1}-m(j_{1}+1)}\gamma _{(\infty
)}^{p_{3}-m(j_{3}+1)}\prod\nolimits_{r=1}^{s}\beta
_{(w_{r})}^{m}\right\rangle _{\lambda =0}\left\langle \bar{\gamma}_{(0)}^{%
\bar{p}_{1}-m(j_{1}+1)}\bar{\gamma}_{(\infty )}^{\bar{p}_{3}-m(j_{3}+1)}%
\prod\nolimits_{r=1}^{s}\bar{\beta}_{(\bar{w}_{r})}^{m}\right\rangle
_{\lambda =0} \ ,
\end{equation}%
where the subscript $\lambda =0$ refers to the fact that these correlation
functions are defined in terms of the free action $S[\lambda =0]$. Standard
free field techniques enable us to write%
\begin{equation*}
W_{(j_{1},j_{2},j_{3}|p_{1},m(1+j_{2}),p_{3})}^{(m)}=\frac{\Gamma
(1-m(j_{1}+1)+\bar{p}_{1})\Gamma (1-m(j_{3}+1)+p_{3})}{\Gamma
(m(j_{1}+1)-p_{1})\Gamma (m(j_{3}+1)-\bar{p}_{3})}\times
\end{equation*}%
\begin{equation}
\times \int
\prod\nolimits_{r=1}^{s}d^{2}w_{r}\prod%
\nolimits_{r=1}^{s}|w_{r}|^{-4b^{2}(j_{1}+1)-2m}|1-w_{r}|^{-4b^{2}(j_{2}+1)}%
\prod\nolimits_{r<t\leq s}|w_{r}-w_{t}|^{-4b^{2}} \ .  \label{INT}
\end{equation}

The $\Gamma $-functions in the first line come from the multiplicity factor
when contracting the fields of $\beta $-$\gamma $ system. This contribution
can be obtained as in \cite{GN3} by generalizing the procedure in \cite{BB}.
This yields

\begin{equation}
\left\langle \gamma _{(0)}^{p_{1}-m(j_{1}+1)}\gamma _{(\infty
)}^{p_{3}-m(j_{3}+1)}\prod\nolimits_{r=1}^{s}\beta
_{(w_{r})}^{m}\right\rangle =\lim_{w_{i}^{(t)}\rightarrow w_{i}^{(1)}=w_{i}}%
\mathcal{P}^{-1}\frac{\partial ^{ms}\mathcal{P}}{\partial
w_{1}^{(1)}...\partial w_{1}^{(m)}...\partial w_{s}^{(1)}...\partial
w_{s}^{(m)}}  \label{pes}
\end{equation}%
with 
\begin{equation}
\mathcal{P}=\prod\nolimits_{r=1}^{s}\prod%
\nolimits_{t=1}^{m}(w_{r}^{(t)})^{m(j_{1}+1)-p_{1}}%
\prod_{r<l}(w_{r}^{(t)}-w_{l}^{(t)}) \ .
\end{equation}%
Here, we regulate the correlation function by point-splitting method for the
insertion points of the screening operators, as $\beta
_{(w_{r})}^{m}\rightarrow $ $\prod\nolimits_{t=1}^{m}\beta _{(w_{r}^{(t)})}$%
, taking $ms$ different points as $w_{r}^{(t)},$ with $r=1,...,s$ and $%
t=1,...,m$, and then taking the coincidence limit $w_{r}^{(t)}\rightarrow
w_{r}^{(t-1)}\rightarrow $ $...w_{r}^{(2)}\rightarrow w_{r}^{(1)}=w_{r}$.
Accordingly, for a particular case $p_{i}=\bar{p}_{i}$, one obtains 
\begin{eqnarray*}
\left| \left\langle \gamma _{(0)}^{p_{1}-m(j_{1}+1)}\gamma _{(\infty
)}^{p_{3}-m(j_{3}+1)}\prod\nolimits_{r=1}^{s}\beta
_{(w_{r})}^{m}\right\rangle \right| ^{2} &=&(-1)^{ms}\gamma
(1-m(j_{1}+1)+p_{1})\times \\
&&\times \gamma (1-m(j_{3}+1)+p_{3})\prod\nolimits_{r=1}^{s}|w_{r}|^{-2m} \ .
\end{eqnarray*}

On the other hand, the generalized Selberg integral in the second line of (%
\ref{INT}) can be computed by using the formula (\ref{ISAH}) of Appendix A,
developed by Dotsenko and Fateev in \cite{DF}. The result takes the form%
\begin{equation*}
C_{(j_{1},j_{2},j_{3}|p_{1},m(1+j_{2}),p_{3})}^{(m)}=b^{-1}\lambda ^{s}\pi
^{s}\Gamma (-s)\Gamma (s+1)\gamma ^{s}\left( 1+b^{2}\right) \delta \left(
\sum\nolimits_{i=1}^{n}j_{i}+n+s-1-b^{-2}(1-m)\right) \times
\end{equation*}%
\begin{equation}
\times \frac{\Gamma (1-m(j_{1}+1)+\bar{p}_{1})\Gamma (1-m(j_{3}+1)+p_{3})}{%
\Gamma (m(j_{1}+1)-p_{1})\Gamma (m(j_{3}+1)-\bar{p}_{3})}\prod_{r=0}^{s-1}%
\frac{\gamma \left( -(r+1)b^{2}\right) \gamma \left( b^{2}\left(
j_{1}-j_{3}+j_{2}+1+r\right) \right) }{\gamma \left(
b^{2}(2j_{1}+2+r)+m\right) \gamma \left( b^{2}(2j_{2}+2+r)\right) } \ .
\label{RES}
\end{equation}

Now, analytic continuation of this expression is needed in order to find the
general result, incorporating also the configurations yielding non-integer $%
s $. Such analytic continuation is done by requiring the residue of the
exact expression evaluated at $s=-2+b^{-2}(1-m)-j_{1}-j_{2}-j_{3}\mathbb{\in
Z}_{\geq 0}$\ to coincide with (\ref{RES}). The analytic continuation yields%
\begin{eqnarray}
C_{(j_{1},j_{2},j_{3}|p_{1},m(1+j_{2}),p_{3})}^{(m)} &=&\left( \pi \lambda
b^{-2}\gamma (b^{2})\right) ^{s}\frac{\Gamma (1-m(j_{1}+1)+\bar{p}%
_{1})\Gamma (1-m(j_{3}+1)+p_{3})}{b\ \Gamma (m(j_{1}+1)-p_{1})\Gamma
(m(j_{3}+1)-\bar{p}_{3})}\times  \notag \\
&&\times \frac{G\left( 1+j_{1}+j_{2}+j_{3}+(m-2)b^{-2}\right) G\left(
j_{1}-j_{2}+j_{3}+(m-1)b^{-2}\right) }{G\left( 2j_{1}+1+(m-1)b^{-2}\right)
G\left( 2j_{3}+1+(m-1)b^{-2}\right) }\times  \notag \\
&&\times \frac{G\left( -j_{1}+j_{2}+j_{3}-b^{-2}\right) G\left(
j_{1}+j_{2}-j_{3}-b^{-2}\right) }{G(-1)G\left( 2j_{2}+1-b^{-2}\right) }\ ,
\label{RFIN}
\end{eqnarray}%
where $s=-2-j_{1}-j_{2}-j_{3}+(1-m)b^{-2}$, and the definition of the $G $%
-function can be found in Appendix A. To derive this expression one can uses
the functional relation 
\begin{equation}
\gamma (-rb^{2})=\frac{\Gamma (-rb^{-2})}{\Gamma (1+rb^{-2})}=\frac{G(r)}{%
G(r-1)} \ .
\end{equation}

Expression (\ref{RFIN}) gives the three-point function for the case\footnote{%
Understanding the kinematic configuration $p_{i}=m(j_{i}+1)$ in general
would require the analysis of the spectrum of the CFT corresponding to
generic values of $m$ and $b$.} $p_{2}=\bar{p}_{2}=m(j_{2}+1)$, for generic
values of $m$ and $b$. Of course, when $m=1$ this expression coincides with
that for the $SL(2,\mathbb{R})_{k}$ WZNW model for the case where one of the
vertex operators is given by the highest-weight representation of $SL(2,%
\mathbb{R})$ (with the identification $k=b^{-2}+2$). In \cite{GN3}, it was
shown how it reproduces the Melin transform of the three-point functions of 
\cite{T1,T2,T3}. More precisely, considering $m=1$ in (\ref{RFIN}) and the
functional relation $G(x)=\gamma (-x)(k-2)^{(2x+1)}G(x+2-k)$, one finds Eqs.
(60)-(62) of \cite{GN3}, which yields the expected result. More remarkably,
the same result is obtained for the case $m=b^{2}=k-2$, as it was discussed
in \cite{GN3}, see Eq. (50) there. As a simpler consistency check, one can
also verify that $m=0$\ in (\ref{RFIN}) yields the correlation function of
three exponential fields $e^{\sqrt{2}\alpha _{i}\varphi (z_{i})}$ of
Liouville field theory, with $\alpha _{i}=b(j_{i}+1)$; namely $%
C_{(j_{1},j_{2},j_{3})}^{(m=0)}=C^{L}(\alpha _{1},\alpha _{2},\alpha _{3})$
by writing $G(x)$ in terms of $\Upsilon (x)$ by using (\ref{Cacatua}) of
Appendix A (upon neglecting $\beta $-$\gamma $ contribution).

The pole structure of (\ref{RFIN}) can be analyzed as follows: the function $%
G(x)$ develops single poles at%
\begin{equation}
x=p+q\text{\ }b^{-2},\qquad x=-(p+1)-(q+1)\text{\ }b^{-2}  \label{RRRT}
\end{equation}%
for any pair of positive integers $p\in \mathbb{Z}_{\geq 0}$ and $q\in 
\mathbb{Z}_{\geq 0}$. This implies that expression (\ref{RFIN}) presents
poles at%
\begin{eqnarray*}
-j_{1}+j_{2}+j_{3} &=&p+(q+1)\text{\ }b^{-2}\qquad
-j_{1}+j_{2}+j_{3}=-(p+1)-q\text{\ }b^{-2} \\
j_{1}+j_{2}-j_{3} &=&p+(q+1)\text{\ }b^{-2}\qquad j_{1}+j_{2}-j_{3}=-(p+1)-q%
\text{\ }b^{-2}
\end{eqnarray*}%
and at%
\begin{eqnarray*}
j_{1}-j_{2}+j_{3} &=&p+(q+1-m)\text{\ }b^{-2},\qquad \qquad
j_{1}-j_{2}+j_{3}=-(p+1)-(q+m)\text{\ }b^{-2} \\
j_{1}+j_{2}+j_{3} &=&p-1+(q+2-m)\text{\ }b^{-2}\qquad
j_{1}+j_{2}+j_{3}=-(p+2)-(q+m-1)\text{\ }b^{-2} \ .
\end{eqnarray*}

It is worth noticing that these pole conditions remain unchanged if one
first performs the changes $m\rightarrow -mb^{2}$, $j_{i}\rightarrow
-b^{-2}j_{i}$, and then replaces $b^{2}$ by $b^{-2}$. This is actually a
manifestation of the level duality under $k-2\rightarrow (k-2)^{-1}$. The
properties of the three-point function under the level duality can be
understood by introducing the dual function $\tilde{G}(x)$, see (\ref{Gtilde}%
) of Appendix A, which presents poles at $x=p+q$\ $b^{2}$ and at $%
x=-(p+1)-(q+1)$\ $b^{2}$, instead of (\ref{RRRT}). Thus, taking into account
the functional relation $\tilde{G}(xb^{2})=b^{2b^{2}x(x+b^{-2}+1)}G(x)$, one
finds that expression (\ref{RFIN}) can be written in terms of the analogous
quantity but for the inverse parameter $b^{-1}$ with an appropriate
redefinition of the spin variables $j_{i}$.

On the other hand, the pole contributions coming from the $\Gamma $%
-functions in the first line of (\ref{RFIN}) are of a different sort, since
these depend on momenta $p_{i}$. The $p_{i}$-dependent pole conditions
depend on the specific power of fields $\gamma $ in the functional form of
the vertex operators (\ref{Vjp}), while the pole conditions written down
above do not depend on those specific powers $p_{i}-m(j_{i}+1)$.

The two-point function can be also obtained from (\ref{RFIN}) by using the
functional relation 
\begin{equation}
\lim_{j_{2}\rightarrow -1}\frac{G\left( -j_{1}+j_{3}+j_{2}-b^{-2}\right)
G\left( j_{1}-j_{3}+j_{2}-b^{-2}\right) }{G(-1)G\left(
2j_{2}+1-b^{-2}\right) }=2\pi b^{-2}\delta (j_{1}-j_{3})\ .
\end{equation}%
In the limit $j_{2}\rightarrow -1$, operator $\Phi
_{j_{2},p_{2},p_{2}}(z_{2})$ approaches the identity operator, and so we can
write 
\begin{equation*}
\left\langle \Phi _{j_{1},p_{1},\bar{p}_{1}}(z_{1})\Phi _{j_{3},p_{3},\bar{p}%
_{3}}(z_{3})\right\rangle _{CFT}\sim |z_{13}|^{-2\Delta _{j_{1}}}\left( \pi
\lambda b^{2}\gamma (b^{2})\right) ^{s}\gamma (-s)\gamma (1+b^{2}s)\ \delta
(j_{1}-j_{3})\times
\end{equation*}%
\begin{equation}
\times \frac{\Gamma (1-m(j_{1}+1)+\bar{p}_{1})\Gamma
(mb^{-2}(1-m)-mj_{1}-p_{1})}{\Gamma (m(j_{1}+1)-p_{1})\Gamma
(1+mb^{-2}(m-1)+mj_{1}+\bar{p}_{1})}\ ,  \label{tpf}
\end{equation}%
where now $s=-1-2j_{1}+b^{-2}(1-m)$, and, according to (\ref{ZDOS}), we have 
$p_{3}=(mb^{-2}-1)(1-m)-p_{1}$. The limit $j_{2}\rightarrow -1$ of the
three-point function is known to agree with the two-point function up to a $%
b $-dependent ($j$-independent) factor \cite{BB,GN3,Z}. Besides, notice that
we could also take the limit $j_{2}\rightarrow b^{-2}(1-m)$ in (\ref{RFIN}),
and this would also yields a two-point function. This is because operator $%
\Phi _{j_{2}=b^{-2}(1-m)}$ also corresponds to a dimension-zero operator
that can be considered as a (Weyl reflected) conjugate representation of the
identity operator. This is analogous to what happens in the $SL(2,\mathbb{R}%
)_{k}$ WZNW model with the operators $\Phi _{j_{2}=0}$ and $\Phi _{j_{2}=-1}$%
; while inserting one of these operators in the three-point function leads
to the reflection coefficient $\sim R^{H}(j_{1})\delta \left(
j_{1}-j_{2}\right) $, inserting the other leads\footnote{%
After the Melin-Fourier transform of expression (\ref{RFIN}) we also have
the same contribution in the $x$-basis as well; see for instance \cite{GN3}
and references therein.} to the \textquotedblleft unreflected term" $\sim
\delta \left( j_{1}+j_{2}+1\right) $. In the general case (namely generic $m$%
), the generalization of $SL(2,\mathbb{R})$ Weyl reflection is given by $%
j\rightarrow -1-j-b^{-2}(m-1)$, which leaves the conformal dimension (\ref%
{ConfDim}) unchanged. Thus, the limit $j_{2}\rightarrow b^{-2}(1-m)$ gives $%
s=0$ and consequently yields the expression%
\begin{equation}
\left\langle \Phi _{j_{1},p_{1},\bar{p}_{1}}(z_{1})\Phi _{j_{3},p_{3},\bar{p}%
_{3}}(z_{3})\right\rangle _{CFT}\sim |z_{13}|^{-2\Delta _{j_{1}}}\ \delta
(j_{1}+j_{3}+1+b^{-2}(m-1))\ .  \label{tpf2}
\end{equation}%
This expression can be obtained from (\ref{tpf}) by replacing $%
j_{3}\rightarrow -1-j_{3}-b^{-2}(m-1)$ before evaluating the $\delta $%
-function $\delta (j_{1}-j_{3})$. Summarizing, (\ref{tpf}) and (\ref{tpf2})
give all the contributions to the two-point function.

To conclude this section, let us comment on an alternative free field
representation. Presumably, some of the CFTs corresponding to certain values
of $m$ and $b$ could also be realized by coupling Liouville theory in a
non-minimal way to a free boson $\eta $ plus a linear dilaton theory for a
field $\chi $ with background charge $Q_{\chi }^{(m)}=$ $%
(mb^{-2}(m-2)-2m)^{1/2}$, and then perturbing the whole theory by
introducing an operator $\Phi _{\kappa }=e^{\kappa \chi }$ with $\kappa $
satisfying $\kappa (Q_{\chi }-\kappa )+\Delta _{\alpha =-m/2b}=-\kappa
^{2}+Q_{\chi }^{(m)}\kappa -m/2-mb^{-2}(1+m/2)/2=1$, and eventually dressing 
$\Phi _{\kappa }$ with the appropriate Liouville field $V_{-m/2b}=e^{-m%
\varphi /\sqrt{2}b}$. For the case $m=1$, this was done in \cite{G2005}. For
this case, as well as for $m=b^{2}$, the background charge corresponds to $%
Q_{\chi }^{(1)}=Q_{\chi }^{(b^{2})}=-i\sqrt{k}$. In particular, this
realization reproduces the correct value of the central charge as $%
c_{_{CFT}}=c_{L}+2+6(Q_{\chi }^{(m)})^{2}=3+6(b+b^{-1}(1-m))^{2}$.
Correlation functions in terms of such a Coulomb-gas representation could be
computed by similar means.

\section{Relation to Hamiltonian reduction}

In this section, we discuss the relation between Hamiltonian reduction and
SRT formula. Eventually, this relation gives us a concrete realization of
the Langlands duality in correlation functions of $H_{3}^{+}$ WZNW model.

\subsection{Reviewing Drinfeld-Sokolov Hamiltonian reduction}

Hamiltonian reduction yields a way of reducing $SL(2,\mathbb{R})_{k}$ WZNW
model to LFT. This can be regraded as a reduction of the degrees of freedom
of WZNW by imposing constraints on $SL(2,\mathbb{R})$ momenta. This results
in LFT that governs the remnant dynamics.

The procedure is as follows: first, impose the gauge%
\begin{equation}
J^{+}(z)=k \ ,  \label{Constraint}
\end{equation}%
and its anti-holomorphic partner. This is implemented by means of the BRST
method: we introduce a $b$-$c$ ghost system with central charge $c_{g}=-2 $
and define the BRST charge as\footnote{%
See also \cite{NP} for a very interesting discussion.}%
\begin{equation*}
Q_{_{BRST}}=\frac{1}{2\pi i}\oint dw\left( J^{+}(w)-k\right) c(w) \ .
\end{equation*}%
This is analogous to the BRST\ implementation of the $SL(2,\mathbb{R}%
)_{k}/U(1)$ WZNW coset construction, see for instance \cite{BB}\ and
references therein.

Current $J^{+}(z)$ originally corresponds to a primary field of conformal
dimension $+1$ with respect to the Sugawara stress tensor. Therefore, in
order to impose (\ref{Constraint}) in a coordinate invariant way, one has to
perform a change in the stress tensor to turn $J^{+}$ into a dimension-zero
operator. This change in the stress tensor is usually referred to as an 
\textit{improvement} or \textit{topological twist}, which is defined by the
shifting%
\begin{equation*}
T(z)\rightarrow \hat{T}(z)=T(z)-\partial J^{3}(z) \ .
\end{equation*}%
Taking into account Wakimoto representation%
\begin{eqnarray*}
T(z) &=&-\beta (z)\partial \gamma (z)-\frac{1}{2}\left( \partial \phi
\right) ^{2}-\frac{1}{\sqrt{2k-4}}\partial ^{2}\phi (z) \ ,\quad \\
J^{3}(z) &=&-\beta (z)\gamma (z)+\sqrt{\frac{k-2}{2}}\partial \phi (z)\ ,
\end{eqnarray*}%
the \textit{improved} stress tensor takes the form%
\begin{equation}
\hat{T}(z)=-\frac{1}{2}\left( \partial \phi \right) ^{2}-\frac{1}{\sqrt{%
2(k-2)}}\partial ^{2}\phi (z)+\partial \beta (z)\gamma (z)-\sqrt{\frac{k-2}{2%
}}\partial ^{2}\phi (z) \ .  \label{ININ}
\end{equation}

One can verify that the OPE $\hat{T}(z)\beta (w)$ is consistent with
treating $J^{+}(z)$ as a zero-dimension field 
\begin{equation*}
\hat{T}(z)\beta (w)\sim \frac{\partial \beta (z)}{(z-w)}+... \ .
\end{equation*}

Besides, from the Wakimoto realization $J^{+}(z)=\beta (z)$ the
implementation of constraint $J^{+}(z)=k$ yields $\partial \beta (z)=0$ in (%
\ref{ININ}); namely%
\begin{equation}
\hat{T}(z)=-\frac{1}{2}\left( \partial \phi \right) ^{2}+\frac{Q}{\sqrt{2}}%
\partial ^{2}\phi (z) \ ,  \label{HATT}
\end{equation}%
where $Q=b+b^{-1}$ and $b^{-2}=k-2$. This actually corresponds to the
Liouville stress-tensor. Thus, we see how the WZNW theory reduces to LFT by
implementing constraint (\ref{Constraint}). Computing the OPE $\hat{T}(z)%
\hat{T}(w)$ one also verifies that the central charge of LFT is given by%
\begin{equation*}
\hat{c}=1+6Q^{2}=c_{_{SL(2)}}+6k-2 \ ,
\end{equation*}%
where $c_{_{SL(2)}}=3+6/(k-2)$ is the central charge of the $SL(2,\mathbb{R}%
)_{k}$ WZNW theory.

Now, we should specify how the spectrum of WZNW theory relates to the
spectrum of LFT, which are represented by the exponential primary fields $%
V_{a}(z)=e^{\sqrt{2}\alpha \phi (z)}$. First, recall the formula for the
conformal dimension of these fields: 
\begin{equation}
\Delta _{\alpha }=\alpha (Q-\alpha ) \ .  \label{YucatanA}
\end{equation}

On the other hand, in the WZNW side it is convenient to focus on the fields
belonging to highest-weight representations of $SL(2,\mathbb{R})$, namely
those fields satisfying $j=-p=-\bar{p}$ (or its Weyl reflected counterpart $%
j+1=p=\bar{p}$). These are primary fields with respect to the improved
stress-tensor $\hat{T}(z)$, as it can be checked by computing the OPE $\hat{T%
}(z)\Phi _{j,p,\bar{p}}(w)$, whose conformal dimension is%
\begin{equation}
\hat{\Delta}_{j}=-\frac{j(j+1)}{k-2}-j=-bj(Q-(-bj))\ .  \label{YucatanB}
\end{equation}

Thus, comparing (\ref{YucatanA}) with (\ref{YucatanB}) we see that it is
natural to identify the Liouville momentum $\alpha $ and the WZNW momentum $%
j $ by the simple relation%
\begin{equation}
\alpha =-bj\ ,  \label{OTAU1}
\end{equation}%
or its Weyl reflected counterpart $\alpha =b\left( j+1\right) $, depending
on the conventions. Namely, Hamiltonian reduction induces the following
identification $V_{\alpha =-bj}(z)\leftrightarrow \Phi _{j,-j,-j}(z)$
between vertex operators of both theories. According to this correspondence,
we expect that the Hamiltonian reduction would be realized at the level of
correlation functions through the form%
\begin{equation}
\left\langle \prod\nolimits_{i=1}^{n}V_{-bj_{i}}(z_{i})\right\rangle
_{LFT}\sim f(b)\left\langle \prod\nolimits_{i=1}^{n}\Phi
_{j_{i}}^{(0)}(z_{i})\right\rangle _{H_{3}^{+}}  \label{punchaA}
\end{equation}%
with the notation 
\begin{equation*}
\Phi _{j}^{(0)}(z)=\mathcal{N}(j,b)\ \Phi _{j,p=-j,\bar{p}=-j}(z)\ ,
\end{equation*}%
where $p_{i}+j_{i}=\bar{p}_{i}+j_{i}=0$, $f(b)$ is some $b$-dependent
numerical factor, and $\mathcal{N}(j,b)$ is some normalization of the vertex
operators. In the following, we will discuss a realization of (\ref{punchaA}%
) in terms of SRT correspondence.

\subsection{Implementing the reduction as the $\protect\mu _{i}\rightarrow 0$
limit}

We would like to attempt interpretation of the SRT formula from the
viewpoint of the quantum DS Hamiltonian reduction. As discussed above, the
conventional way of implementing the reduction is by imposing the gauge
condition (\ref{Constraint}), which in the Wakimoto representation reads $%
\beta =k$. On the other hand, in the path integral formulation discussed in
section 2.2, integration over $\gamma $ has given the condition 
\begin{equation*}
\bar{\partial}\beta (w)=2\pi \sum_{i=1}^{n}\mu _{i}\delta (w-z_{i})\ ,
\end{equation*}%
or, equivalently, 
\begin{equation}
\beta (w)=\text{const}+\sum_{i=1}^{n}\frac{\mu _{i}}{w-z_{i}}\ .
\label{betaff}
\end{equation}%
As a 1-form, the constant should vanish in principle, but after the \textit{%
improvement} of the energy momentum tensor, the constant term would be
allowed\footnote{%
As mentioned, this improvement can be obtained by shifting the Sugawara
stress-tensor $T(z)$ as $T(z)\rightarrow T(z)-\partial J^{3}(z)$.
Alternatively, one can add a piece $\bar{\omega}J$ in the action, where $%
\omega $ is a worldsheet connection, which leads to the equation of motion $(%
\bar{\partial}+\bar{\omega})\beta (w)=2\pi \sum_{i=1}^{n}\mu _{i}\delta
(w-z_{i})$.}.

Now, let us take the limit $\mu _{i}\rightarrow 0$ (while keeping $%
\sum_{i=1}^{n}\mu _{i}=0$) in the (dual) SRT formula. More precisely, we
first take $\mu _{i}\rightarrow 0$ for $i=2,3,\cdots n-1$, and then we take
the further limit $\mu _{1}=-\mu _{n}\rightarrow 0$. This limit is in
harmony with the spirit of the Hamiltonian reduction because \eqref{betaff}
suggests that in order to fix $\beta $ to be a constant number, $\mu
_{i}\rightarrow 0$ limit seems unavoidable. In this limit, the parameter $%
y_{i}$ appearing in the SRT formula becomes $y_{i}\rightarrow z_{i+1}$ for $%
i=1\cdots n-2$ through the relation 
\begin{equation*}
\mu _{i}=u\frac{\prod_{j=1}^{n-2}(z_{i}-y_{j})}{\prod_{j\neq i}(z_{i}-z_{j})}%
\ .
\end{equation*}

The crucial observation is that in this limit we can essentially remove $n-2$
extra vertex operators\footnote{%
The other two vertex operators seem special as in the $H_{3}^{+}$ side. In
the $H_{3}^{+}$ side, we do not necessarily take $\mu \rightarrow 0$ limit
for two vertex operators. In the Liouville side, the relation between $j$
and $\alpha $ is not modified. These vertex operators can be put at $0$ and $%
\infty $ and can be regarded as \textit{in} and \textit{out} vacua instead.}
in the Liouville side of the (dual) SRT formula \eqref{UPP}. This is because
we take the limit $y_{i}\rightarrow z_{i+1}$ and degenerate field $V_{-%
\tilde{b}/2}$ collide with $n-2$ fields $V_{\alpha }$ in the Liouville
correlation functions. Since $V_{-\tilde{b}/2}$ is a degenerate field, it
takes the very simple OPE 
\begin{equation*}
V_{\alpha }(z)V_{-\tilde{b}/2}(w)=\frac{V_{\alpha -\tilde{b}/2}(w)}{%
|z-w|^{-2\alpha \tilde{b}}}+\tilde{C}_{-}^{L}(\alpha )\frac{V_{\alpha +%
\tilde{b}/2}(w)}{|z-w|^{-2(Q-\alpha )\tilde{b}}}\ +(\text{descendants})\ .
\end{equation*}%
Thus, for $\alpha <Q/2$, which is in the so-called Seiberg bound
(corresponding to $j<-1/2$), only the first term in the OPE survives because
it dominates over the second term in the $w\rightarrow z$ limit, and the
Liouville side of the SRT formula then is given by the $n$-point correlation
function 
\begin{equation*}
\left\langle V_{\alpha _{1}}(z_{1})\prod\nolimits_{i=2}^{n-1}V_{\tilde{\alpha%
}_{i}}(z_{i})V_{\alpha _{n}}(z_{n})\right\rangle _{LFT}\ ,
\end{equation*}%
where $\alpha _{i}=\tilde{b}^{-1}(j_{i}+1+\tilde{b}^{2}/2)=\tilde{b}%
^{-1}(j_{i}+k/2)$ and $\tilde{\alpha}_{i}=\tilde{b}^{-1}(j_{i}+1)$. The
factor $\Theta _{n}$ becomes singular in the limit $y_{i}\rightarrow z_{i+1}$%
, but it is not difficult to remove this singularity, and one can see (up to
some normalization constant $\mathcal{C}$) that 
\begin{equation}
\left\langle \prod\nolimits_{i=1}^{n}\Phi _{j_{i}}(0|z_{i})\right\rangle =%
\mathcal{C}\ |\tilde{\Theta}_{n}|^{2}\left\langle V_{\alpha
_{1}}(z_{1})\prod\nolimits_{i=2}^{n-1}V_{\tilde{\alpha}_{i}}(z_{i})V_{\alpha
_{n}}(z_{n})\right\rangle _{LFT}\ ,  \label{YuFormula}
\end{equation}%
where $\tilde{\Theta}_{n}=\prod_{1<i<j<n-1}(z_{i}-z_{j})^{2\tilde{b}^{2}}$
is the regulated version of $\Theta _{n}$ in the SRT formula. A similar
singular normalization factor appeared in the earlier attempt of Hamiltonian
reduction by setting $x_{i}=z_{i}$ \cite{Rasmussen2}.

Notice that the relation between $\tilde{\alpha}_{i}$ and $j_{i}$ in (\ref%
{YuFormula}) agrees with the Weyl reflected version of (\ref{OTAU1}), namely
performing $j_{i}\rightarrow -1-j_{i}$ there. Besides, the relation between $%
\alpha _{i}$ and $j_{i}$ corresponds to performing the change $%
j_{i}\rightarrow -k/2-j_{i}$ in (\ref{OTAU1}). We further discuss this point
in Appendix B and argue this is consistent with what is expected from
Hamiltonian reduction. We propose that the formula (\ref{YuFormula}) yields
the Hamiltonian reduction interpretation of the SRT formula.

\subsection{Realizing Langlands level duality}

Now, let us make a remark on level duality in WZNW theory. As we did in (\ref%
{UPP}), it is possible to derive a dual version of formula of (\ref%
{YuFormula}) by replacing $\tilde{b}$ and $b^{-1}$. Since LFT does not have
any extra insertion in the particular limit we have considered (i.e. $\mu
_{i}\rightarrow 0$), by equating the two expressions (the one in the SRT
formula and the other in the dual SRT formula) with the crucial
identification $b=\tilde{b}$, then we find the following surprising identity%
\footnote{%
Strictly speaking, since the cosmological constant of the LFT is different
between the original formula and the dual formula, we have to adjust the
screening parameter in $H_3^+$ model ($\lambda$ in the Wakimoto realization)
so that the both sides show the same scaling behavior.} 
\begin{equation}
\left\langle \prod\nolimits_{i=1}^{n}\Phi _{j_{i}}(0|z_{i})\right\rangle
_{k}=\tilde{\mathcal{C}}\ \left\langle \prod\nolimits_{i=1}^{n}\Phi _{\tilde{%
j}_{i}}(0|z_{i})\right\rangle _{\tilde{k}}\ ,  \label{zorroZ}
\end{equation}%
where the levels of the WZNW model in the both side of this expression are
related though the Langlands duality 
\begin{equation}
\tilde{k}-2=(k-2)^{-1},
\end{equation}%
and we introduced a numerical coefficient $\tilde{\mathcal{C}}$ that
regularizes the $\mu _{i}\rightarrow 0$ limit. The spin of the vertex
operators on each side obey the relation $\tilde{j}_{i}=(j+1)/(k-2)-1$ for $%
i=1,n$, and $\tilde{j}_{i}=\left( j+1\right) /(k-2)-1/2-(k-2)^{-1}/2$ for $%
i=2,\cdots n-1$. One can explicitly check that this identity is true for the
two-point functions and the three-point functions (see Appendix A for useful
identities such as \eqref{Gtilde}).

This identity can be regarded as a manifestation of the quantum Langlands
duality at the level of correlation function. In physical applications,
it shows a strong/weak coupling duality between different CFTs even with
different central charges. In the context of string theory, it might relate
the scattering amplitudes of the completely different string
compactifications even with different target space dimensionality. For
example, if we embed the $SL(2,\mathbb{R})/U(1)$ coset model in the
superstring compactification, the supersymmetric version of the Langlands
duality is given by $\hat{k}\rightarrow 1/\hat{k}$, where $\hat{k}=k-2$ is
the supersymmetric level of the current algebra. This gives a non-trivial
duality between the scattering amplitudes in the two-dimensional black hole
for $\hat{k}=1/2$ and those in the $A_{1}$ type singularity (Eguchi-Hanson
space) for $\hat{k}=2$. From the viewpoint of the AdS$_{3}$/CFT$_{2}$
correspondence, it predicts a strong-weak duality in the boundary conformal
field theory as well, whose origin is rather mysterious.

~From the mathematical point of view, an identity like (\ref{zorroZ}) could
give a clue to understand the quantum version of the geometric Langlands
correspondence, which is yet to be fully formulated in precise mathematical
language. See e.g. \cite{Stoyanovsky:2006mj} for an interesting discussion
on an attempt.

\section{Discussion and outlook}

In this paper, we have investigated the interrelationship among the
following three notions: the Langlands level duality $k-2\rightarrow \tilde{k%
}-2=1/(k-2)$, the SRT $H_{3}^{+}-$Liouville correspondence, and the DS
Hamiltonian reduction. First, we have derived a dual version of $H_{3}^{+}-$%
Liouville correspondence formula (\ref{UPP}) induced by the Liouville
self-duality under $b\rightarrow b^{-1}$. We have also discussed how the dual
formula can be interpreted as a particular case of the non-rational CFTs
recently proposed in \cite{R}. By using the free field realization, we have
confirmed that the $H_{3}^{+}$ WZNW model is actually double-represented
within the space of parameters ($m,b$). The free field techniques have also
enabled us to compute three-point functions for these non-rational CFTs.

The dual formula (\ref{UPP}), together with the original formula of \cite{RT}%
, show how Langlands level duality of $H_{3}^{+}$ WZNW model manifests itself at
the level of $n$-point correlation functions. This is particularly realized
in (\ref{zorroZ}) (see also (\ref{zorroZZ}) of Appendix B). We have argued
how such equations can be regarded as a realization of Hamiltonian reduction
at the level of correlation functions. More precisely, we have proposed its
realization in terms of a limit of the (dual) SRT relation. This is
represented in (\ref{YuFormula}) derived in the limit $\mu _{i}\rightarrow 0$%
, which corresponds to the Hamiltonian reduction in $\mu$ basis.

Studying the relation between the SRT $H_{3}^{+}-$Liouville correspondence
and the DS Hamiltonian reduction could be relevant in the context of the
geometric Langlands correspondence. In fact, (\ref{YuFormula}) (see also (%
\ref{punchaB}) in Appendix B) can be seen as a quantum non-chiral version of
the geometric Langlands correspondence, in the sense that this formula
selects precise basis of the WZNW $n$-point correlation functions that admit
to be expressed in terms of $n$-point correlation functions of the Virasoro
algebra. We emphasize that the advantage of (\ref{YuFormula}) (see also (\ref%
{punchaB}) in Appendix B) over the other expressions is that it provides a
map between $n$-point functions in both sides, without involving additional
degenerate fields.

Last but not least, we would like to advocate that understanding of the
precise relation between the quantum DS Hamiltonian reduction and the SRT $%
H_{3}^{+}-$Liouville correspondence would be important for its higher rank
generalization. It is commonly believed that an analogous correspondence
should exist between the $SL(N,\mathbb{R})_{k}$ WZNW model and the affine
Toda\footnote{%
Recently, \cite{Fateev:2007ab} computed certain classes of correlation
functions in conformal Toda field theories.} field theory (corresponding to
SRT formula for $N=2$). Since the quantum version of the Hamiltonian
reduction admits a generalization to $N>2$ \cite{Bershadsky:1989mf}, this
approach could be a natural way to tackle this open question to find its
feasible extensions in higher ranks.

\subsection*{Acknowledgement}

G.G. thanks M. Porrati and S. Ribault for interesting comments. The work of
G.G. is supported by UBA, CONICET and ANPCyT, through grants UBACyT X861,
PIP6160, PICT34557. The research of Y.~N. is supported in part by NSF grant
PHY-0555662 and the UC Berkeley Center for Theoretical Physics.

\appendix

\section{Special functions}

Here we summarize some useful formulae on special functions.

The function $\Upsilon (x)$ was introduced by Zamolodchikov and
Zamolodchikov in Ref. \cite{ZZ}, and it is defined by 
\begin{equation}
\Upsilon (x)=\exp \left( \int_{0}^{\infty }\frac{dt}{t}\left[ \left( \frac{Q%
}{2}-x\right) ^{2}e^{-t}-\frac{\sinh ^{2}\left( \frac{Q}{2}-x\right) \frac{t%
}{2}}{\sinh \frac{bt}{2}\sinh \frac{t}{2b}}\right] \right)  \label{Upa}
\end{equation}%
for $0<\mathbb{R}e(x)<Q$, and by its analytic continuation outside the
strip. It satisfies the shift equations 
\begin{equation*}
\Upsilon (x+b)=\gamma (bx)b^{1-2x}\Upsilon (x)\ ,\ \ \Upsilon
(x+b^{-1})=\gamma (xb^{-1})b^{2xb^{-1}-1}\Upsilon (x)\ ,
\end{equation*}%
where $\gamma (x)=\Gamma (x)/\Gamma (1-x)$, which obeys $\gamma (x)\gamma
(1-x)=1$, $\gamma (x)\gamma (-x)=-x^{-2}$. Notice also that definition (\ref%
{Upa}) is invariant under $b\rightarrow b^{-1},$ and this fact yields
further interesting functional relations.

On the other hand, the special function $G(x)$ is defined in terms of $%
\Upsilon (x)$ by 
\begin{equation}
G(x)=\Upsilon ^{-1}(-bx)b^{-b^{2}x^{2}-(b^{2}+1)x}\ ,  \label{Cacatua}
\end{equation}%
and consequently satisfies the shift equations 
\begin{equation}
G(x)=\gamma (-x)b^{-2(2x+1)}G(x-b^{-2})\ ,\ \ G(x+1)=\gamma
(-(1+x)b^{2})G(x)\ .
\end{equation}

Function $G(x)$ can be also defined in terms of Barnes' $\Gamma _{2}$%
-function, as follows%
\begin{equation}
G(x)=b^{x(xb^{2}-1-b^{2})}\Gamma _{2}(-x|1,b^{-2})\Gamma
_{2}(b^{-2}-1+x|1,b^{-2}) \ ,  \label{Tomato}
\end{equation}%
with%
\begin{equation*}
\log \Gamma _{2}(x|1,y)=\lim_{\varepsilon \rightarrow 0}\partial
_{\varepsilon }\left( \sum\nolimits_{n,m}(x+n+my)^{-\varepsilon
}-\sum\nolimits_{n,m}(n+my)^{-\varepsilon }\right) \ ,
\end{equation*}%
where the first sum runs over positive integers $n\in \mathbb{Z}_{\geq 0}$
and $m\in \mathbb{Z}_{\geq 0}$, while the second sum excludes the step
\thinspace $n=m=0$. Function $G(x)$ develops single poles at 
\begin{equation*}
x=n+mb^{-2}\ ,\quad x=-(1+n)-(m+1)b^{-2} \ .
\end{equation*}

It is also useful to introduce the \textit{dual} function \thinspace $\tilde{%
G}(x)$ that is defined as in (\ref{Tomato}) by replacing $b\rightarrow
b^{-1} $. A manifestation of the invariance of (\ref{Upa}) under $%
b\rightarrow b^{-1} $ is the following identity%
\begin{equation}
G(x)=(k-2)^{(x^{2}+(k-1)x)/(k-2)}\ \tilde{G}(x/(k-2))\ ,  \label{Gtilde}
\end{equation}%
where $b^{-2}=k-2$. Relation (\ref{Gtilde}) can be seen from (\ref{Cacatua}).

In the main text, we have also used the following integral formula, which is
known as Dotsenko-Fateev integral 
\begin{equation*}
\frac{1}{m!}\int d^{2}z_{i}\prod_{i=1}^{m}|z_{i}|^{2\alpha
}|1-z_{i}|^{2\beta }\prod_{i<j}^{m}|z_{i}-z_{j}|^{4\rho }=\pi ^{m}(\gamma
(1-\rho ))^{m}\times
\end{equation*}%
\begin{equation}
\times \prod_{i=1}^{m}\gamma (i\rho )\gamma (1+\alpha +(i-1)\rho )\gamma
(1+\beta +(i-1)\rho )\gamma (-1-\alpha -\beta -(m-2+i)\rho )\ .  \label{ISAH}
\end{equation}

\section{Spectral flowed correlation functions}

In this appendix, we discuss an alternative way of realizing the Hamiltonian
reduction at the level of correlation functions in terms of SRT formula. In
particular, we will study the WZNW correlation functions that involve
spectral flowed states. That is, we will consider the correlation functions
that violate the so-called winding number, as defined in \cite{R2}.

As mentioned in the introduction, the SRT formula of \cite{RT} was
generalized in \cite{R2} to the case with spectral flowed (winding) states
of $SL(2,\mathbb{R})_{k}$ WZNW theory. The general result states that $n$%
-point functions of the $H_{3}^{+}$ WZNW theory that violate the winding
conservation in $\Delta \omega $ units is given by a $2n-2-\Delta \omega $%
-point functions in LFT (where $n-2-\Delta \omega $ Liouville vertex
operators correspond degenerate fields $V_{-1/2b}$). Explicitly, we have 
\begin{equation*}
\left\langle \prod\nolimits_{i=1}^{n}\Phi _{j_{i},p_{i},\bar{p}_{i}}^{\omega
_{i}}(z_{i})\right\rangle _{H_{3}^{+}}=\frac{2\pi ^{3-2n}bc_{k}^{\Delta
\omega }}{\Gamma (n-1-\Delta \omega )}\prod\nolimits_{i=1}^{n}\frac{\Gamma
(-j_{i}+p_{i})}{\Gamma (j_{i}+1-\bar{p}_{i})}\times
\end{equation*}%
\begin{eqnarray}
&&\times \prod\nolimits_{1\leq l<t\leq n}\left( z_{l}-z_{t}\right) ^{\frac{k%
}{2}-\frac{k}{2}\omega _{l}\omega _{t}+\omega _{l}p_{t}+p_{l}\omega
_{t}+p_{l}+p_{t}}\left( \bar{z}_{l}-\bar{z}_{t}\right) ^{\frac{k}{2}-\frac{k%
}{2}\omega _{l}\omega _{t}+\omega _{l}\bar{p}_{t}+\bar{p}_{l}\omega _{t}+%
\bar{p}_{l}+\bar{p}_{t}}\times  \notag \\
&&\times \int \prod\nolimits_{t=1}^{n-2-\Delta \omega }d^{2}y_{t}\quad
\prod\nolimits_{l=1}^{n}\prod\nolimits_{t=1}^{n-2-\Delta \omega }\left(
z_{l}-y_{t}\right) ^{\frac{k}{2}-p_{l}}\left( \bar{z}_{l}-\bar{y}_{t}\right)
^{\frac{k}{2}-\bar{p}_{l}}\times  \notag \\
&&\times \prod\nolimits_{1\leq a<b\leq n-2-\Delta \omega
}|y_{a}-y_{b}|^{k}\left\langle \prod\nolimits_{i=1}^{n}V_{\alpha
_{i}}(z_{i})\prod\nolimits_{t=1}^{n-2-\Delta \omega }V_{-\frac{1}{2b}%
}(y_{t})\right\rangle _{LFT}\ ,  \label{YYYYT}
\end{eqnarray}%
where $b^{-2}=k-2$, $\sum\nolimits_{i=1}^{n}\omega _{i}=-\Delta \omega $, $%
\alpha _{i}=b(j_{i}+k/2)$, $\sum\nolimits_{i=1}^{n}p_{i}=\sum%
\nolimits_{i=1}^{n}\bar{p}_{i}=-k\Delta \omega /2$, and $c_{k}$ is a $k $%
-dependent factor; see \cite{R2} for details. Expression (\ref{YYYYT}) is
the most general version of the $H_{3}^{+}-$Liouville correspondence
involving spectral flowed states $\Phi _{j,p,\bar{p}}^{\omega }(z)$ of the
WZNW theory. In particular, for $\Delta \omega =0$ one recovers the SRT
relation between $n$-point WZNW functions and $2n-2$- point Liouville
functions.

On the other hand, in the case of maximally violating amplitudes (i.e. $%
\Delta \omega =n-2$), all the degenerate field $V_{-1/2b}$ in (\ref{YYYYT})
disappear and the formula actually yields the correspondence between $n$%
-point WZNW functions and $n$-point LFT\ functions. In such a case, we have $%
\Delta \omega =-\sum\nolimits_{i=1}^{n}\omega _{i}=n-2$, and (\ref{YYYYT})
takes the form%
\begin{equation*}
\left\langle \prod\nolimits_{i=1}^{n}\Phi _{j_{i},p_{i},\bar{p}_{i}}^{\omega
_{i}}(z_{i})\right\rangle _{H_{3}^{+}}=2\pi
^{3-2n}bc_{k}^{n-2}\prod\nolimits_{i=1}^{n}\frac{\Gamma (-j_{i}+p_{i})}{%
\Gamma (j_{i}+1-\bar{p}_{i})}\left\langle \prod\nolimits_{i=1}^{n}V_{\alpha
_{i}}(z_{i})\right\rangle _{LFT}\times
\end{equation*}%
\begin{equation}
\times \prod\nolimits_{1\leq l<t\leq n}\left( z_{l}-z_{t}\right) ^{\frac{k}{2%
}-\frac{k}{2}\omega _{l}\omega _{t}+\omega _{l}p_{t}+p_{l}\omega
_{t}+p_{l}+p_{t}}\left( \bar{z}_{l}-\bar{z}_{t}\right) ^{\frac{k}{2}-\frac{k%
}{2}\omega _{l}\omega _{t}+\omega _{l}\bar{p}_{t}+\bar{p}_{l}\omega _{t}+%
\bar{p}_{l}+\bar{p}_{t}} \ ,  \label{LaRibault2}
\end{equation}%
where $\sum\nolimits_{i=1}^{n}p_{i}=\sum\nolimits_{i=1}^{n}\bar{p}%
_{i}=k(n-2)/2$. Notice that the notation here is such that spectral flowed
fields $\Phi _{j_{i},p_{i},\bar{p}_{i}}^{\omega _{i}}$ have conformal
dimension given by 
\begin{equation*}
\Delta _{j}^{\omega ,p}=-b^{2}j(j+1)+p\omega -k\omega ^{2}/4 \ .
\end{equation*}

Now, let us investigate two particular cases of (\ref{LaRibault2}), which
are relevant for the Hamiltonian reduction.

\textbf{Case 1}: Firstly, consider two states of spectral flow sector $%
\omega =0$, and $n-2$ states of sector $\omega =-1$. Suppose $\omega
_{1}=\omega _{n}=0$, $p_{1}=p_{1}=j_{i}$, $p_{n}=p_{n}=-j_{n}$, while $%
\omega _{2}=\omega _{3}=...\omega _{n-1}=-1$, $p_{2}=p_{3}=...p_{n-1}=\bar{p}%
_{2}=\bar{p}_{3}=...\bar{p}_{n-1}=-k/2$. In this case, (\ref{LaRibault2})
can be written as follows%
\begin{equation}
\left\langle \prod\nolimits_{i=1}^{n}V_{\alpha _{i}}(z_{i})\right\rangle
_{LFT}=\frac{\pi }{2b}\left| z_{1}-z_{n}\right|
^{2j_{1}+2j_{n}-k}\left\langle \Phi
_{j_{1}}^{(0)}(z_{1})\prod\nolimits_{i=2}^{n-1}\Phi
_{j_{i}}^{(-)}(z_{i})\Phi _{j_{n}}^{(0)}(z_{n})\right\rangle _{H_{3}^{+}} \ ,
\label{KEYY}
\end{equation}%
where we have defined%
\begin{equation*}
\Phi _{j}^{(0)}(z)=\frac{1}{\gamma (-2j)}\Phi _{j,-j,-j}^{\omega
=0}(z),\quad \Phi _{j}^{(-)}(z)=\frac{c_{k}\pi ^{2}}{\gamma (k/2-j)}\Phi
_{j,-\frac{k}{2},-\frac{k}{2}}^{\omega =-1}(z) \ .
\end{equation*}

Expression (\ref{KEYY}) is certainly similar to (\ref{YuFormula}). As in (%
\ref{YuFormula}), it would be convenient to fix the inserting points as $%
z_{1}=0$ and $z_{n}=\infty $ by using projective invariance. This would make
the overall factor $\left| z_{1}-z_{n}\right| ^{2j_{1}+2j_{n}-k}$ to
disappear, yielding a correspondence between $n$-point functions of both
theories.

\textbf{Case 2}: Secondly, consider one state of the spectral flow sector $%
\omega =+1$, and $n-1$ states of sector $\omega =-1$. Now suppose $\omega
_{1}=+1$, $p_{1}=p_{1}=+k/2$, while $\omega _{2}=\omega _{3}=...\omega
_{n-1}=\omega _{n}=-1$, $p_{2}=p_{3}=...p_{n-1}=p_{n}=\bar{p}_{2}=\bar{p}%
_{3}=...\bar{p}_{n-1}=\bar{p}_{n}=-k/2$. In this case, we find%
\begin{equation}
\left\langle \prod\nolimits_{i=1}^{n}V_{-bj_{i}}(z_{i})\right\rangle _{LFT}=%
\frac{\pi }{2b}\left\langle \Phi _{-\frac{k}{2}-j_{1}}^{(+)}(z_{1})\prod%
\nolimits_{i=2}^{n}\Phi _{-\frac{k}{2}-j_{i}}^{(-)}(z_{i})\right\rangle
_{H_{3}^{+}} \ ,  \label{punchaB}
\end{equation}%
where we have introduced 
\begin{equation*}
\Phi _{-\frac{k}{2}-j}^{(-)}(z)=\frac{c_{k}\pi ^{2}}{\gamma (j+k)}\Phi _{-%
\frac{k}{2}-j,-\frac{k}{2},-\frac{k}{2}}^{-1}(z),\quad \Phi _{-\frac{k}{2}%
-j}^{(+)}(z)=\frac{1}{c_{k}\pi ^{2}\gamma (j)}\Phi _{-\frac{k}{2}-j,+\frac{k%
}{2},+\frac{k}{2}}^{+1}(z) \ ,
\end{equation*}%
and 
\begin{equation}
\alpha =b\left( j+k/2\right) \ .  \label{OTAU2}
\end{equation}

Remarkably, in expression (\ref{punchaB}), all the dependence on $%
|z_{i}-z_{j}|$ dropped out without fixing the insertion points $z_{i}$.
Consequently, this can be actually thought of as a direct correspondence
between $n$-point WZNW correlation functions an $n$-point LFT\ correlation
functions. Operator $\Phi _{-k/2-j_{1}}^{(+)}$ in (\ref{punchaB}) should be
regarded as the one defining the $out$ vacuum state, while operators $\Phi
_{-k/2-j_{i>1}}^{(-)}$ act on the $in$ vacuum creating worldsheet string
states. Also notice that field $\Phi _{-k/2-j}^{(+)}$ has the following
conformal dimension with respect to the stress tensor $T(z)$, 
\begin{equation*}
\Delta _{-k/2-j}^{\omega =-1,p=-k/2}=\frac{(j+k/2)(-j-k/2+1)}{k-2}+\frac{k}{4%
}=-\frac{j(j+1)}{k-2}-j \ ,
\end{equation*}%
which certainly agrees with the formula (\ref{YucatanB}) for the conformal
dimension of fields $\Phi _{j,-j,-j}^{0}$ with respect to the improved
stress tensor $\hat{T}(z)=T(z)-\partial J^{3}(z)$. In turn, we have%
\begin{equation*}
\hat{\Delta}_{j}=\Delta _{-k/2-j}^{\omega =\pm 1,p=\pm k/2} \ .
\end{equation*}

Hence, fields $\Phi _{j,-j,-j}^{0}$ and $\Phi _{-k/2-j,\pm k/2,\pm k/2}^{\pm
1}$ have the same conformal dimension, and these also represent same value
of Liouville momentum $\alpha =-bj$, as in (\ref{OTAU1}). In this sense, we
can associate fields as $\Phi _{j}^{(0)}(z)\leftrightarrow \Phi
_{-k/2-j}^{(\pm )}(z)$. This manifestly shows the parallelism between
realizations (\ref{punchaA}) and (\ref{punchaB}). Moreover, it is also
consistent with the relation between (\ref{OTAU1}) and (\ref{OTAU2}). We
conclude that (\ref{punchaB}) can be thought of as a realization of
Hamiltonian reduction at the level of correlation functions.

To understand the emergence of spectral flowed sector $\omega =-1$ in (\ref%
{punchaB}), one has to take into account that implementing the condition $%
J^{+}=\beta =k$ induces a shifting\footnote{%
G.G. thanks M. Porrati for pointing it out.} of the modes $J_{n}^{\pm ,3}$
of Kac-Moody currents $J_{(z)}^{\pm ,3}=\sum_{n}J_{n}^{\pm ,3}$ $z^{-n-1}$,
and such shifting can be interpreted as a spectral flow transformation with
parameter $\omega =-1$, which yields the flow $J_{n}^{\pm }\rightarrow
J_{n\mp 1}^{\pm }$, $J_{0}^{3}\rightarrow J_{0}^{3}+k/2$. Again, this is
related to the fact that the conformal dimension of a field $\Phi
_{j}^{(0)}\sim \Phi _{j,j,j}^{(\omega =0)}$ with respect to the improved
stress tensor $T(z)-\partial _{z}J^{3}(z)$ agrees with that of a flowed
field $\Phi _{j}^{(-)}\sim \Phi _{-j-k/2,k/2,k/2}^{(\omega =-1)}$ with
respect to $T(z)$.

Let us notice that, as in the case of the $\mu $-basis, it turns out that
performing the change $b\rightarrow \tilde{b}^{-1}$ in (\ref{punchaB})
yields a dual version of such a formula. In fact, we can write down the dual
formula as 
\begin{equation}
\left\langle \prod\nolimits_{t=1}^{n}V_{\alpha _{t}}(z_{t})\right\rangle
_{LFT}\ \ =\frac{\tilde{b}}{2\ \pi ^{3}}\left\langle \Phi
_{j_{1}}^{(+)}(z_{1})\prod\nolimits_{r=2}^{n}\Phi
_{j_{r}}^{(-)}(z_{r})\right\rangle _{H_{3}^{+}} \ ,  \label{IIT}
\end{equation}%
where now $\alpha =\tilde{b}^{-1}(j+1)+b/2$, $\tilde{b}^{2}=k-2$. Then, from
(\ref{punchaB}) and (\ref{IIT}) we obtain the duality relation 
\begin{equation}
\left\langle \Phi _{{j}_{1}}^{(+)}(z_{1})\prod\nolimits_{r=2}^{n}\Phi _{{j}%
_{r}}^{(-)}(z_{r})\right\rangle _{k}=\hat{\mathcal{C}}\ \left\langle \Phi _{%
\tilde{j}_{1}}^{(+)}(z_{1})\prod\nolimits_{r=2}^{n}\Phi _{\tilde{j}%
_{r}}^{(-)}(z_{r})\right\rangle _{\tilde{k}}\ ,  \label{zorroZZ}
\end{equation}%
where $\hat{\mathcal{C}}=(k-2)^{-2}$, and $\tilde{j}=(k-2)\left(
j+1/2\right) -1/2$ and $\tilde{k}-2=(k-2)^{-1}$. This identity again can be
regarded as a manifestation of the quantum Langlands duality at the level of
correlation functions by relating the strongly coupled system with the
weakly coupled system.

To conclude this appendix, let us make a remark on the generalization of the
formula (\ref{YYYYT}) to the case of higher genus correlation functions. In
fact, it would be very interesting to extend the $H_{3}^{+}-$Liouville
correspondence to the case of higher-genus correlation functions involving
spectral flowed sectors (winding sectors of string theory in AdS$_{3}$). An
intriguing result for $SL(2,\mathbb{R})_{k}$ WZNW correlation functions on
the sphere is the existence of an upper bound for the violation of the
winding number conservation. It turns out that winding number conservation
in a given tree-level $n$-point function can be violated up to $n-2$ units%
\footnote{%
This bound can be understood in terms of the $sl(2)_{k}$ symmetry of the
theory; see appendix D of \cite{MO3} for a discussion.}.

In the context of the $H_{3}^{+}-$Liouville correspondence, this upper bound 
$\Delta \omega \leq n-2$ for the winding violation is nicely realized as
follows: according to (\ref{YYYYT}) the $n$-point WZNW correlation functions
that violate the winding conservation in $\Delta \omega $ units are related
to $2n-2-\Delta \omega $-point functions of LFT, where $n-2-\Delta \omega $
Liouville vertex operators represent degenerate fields $V_{-1/2b}$. In other
words, violating the winding conservation in $\Delta \omega $ units on the
WZNW side corresponds to removing $\Delta \omega $ degenerate fields on the
Liouville side of the original formula in \cite{RT}. Then, a natural
question is how such a picture is generalized to the case of higher genus
correlation functions. Presumably, for a genus-$g$ correlation function the
general story remains the same, and one could associate the units of winding
violation to the amount of degenerate Liouville fields $V_{-1/2b}$.
Actually, one is tempted to conjecture that the upper bound for the
violation of the winding number in a genus-$g$ $n$-point correlation
function is given by $\Delta \omega \leq n+2g-2$. On one hand, this is the
amount of Liouville degenerate fields in the case of winding number
conserved amplitudes \cite{HS}; on the other hand, it numerically matches
with the expected number if one thinks the maximally violating genus-$g$
correlation function as factorized in terms of several maximally violating
genus-zero$\ $correlation functions. It would be very interesting to confirm
that such a bound is obeyed.

\bigskip

\end{document}